\documentclass[onecolumn,sort&compress,numbers]{els-mrw} 
\usepackage{amsmath,amssymb,amsfonts,amsthm,makeidx,graphicx}

\usepackage{txfonts}
\usepackage{helvet}

\usepackage{hyperref}
\usepackage{float}

\newcommand{\remu}{\mbox{$R_{e/\mu}$}}
\newcommand{\pienu}{\mbox{$\pi^+ \rightarrow e^+ \nu_e$}}
\newcommand{\pimu}{\mbox{$\pi^+ \rightarrow \mu^+ \nu_\mu$}}
\newcommand{\pimue}{\mbox{$\pi \rightarrow \mu \rightarrow e$}}
\newcommand{\mudecay}{\mbox{$\mu^+ \rightarrow e^+ \nu_e \overline{\nu_\mu}$}}
\newcommand{\pibeta}{\mbox{$\pi^+ \rightarrow \pi^0  e^+ \nu_e$}}

\usepackage{graphicx}

\usepackage{bm}
\usepackage{framed}

\newcommand{\beq}{\begin{equation}}
\newcommand{\eeq}{\end{equation}}
\newcommand{\beqs}{\begin{eqnarray}}
\newcommand{\eeqs}{\end{eqnarray}}
\newcommand{\lsim}{\mathrel{\raisebox{-
.6ex}{$\stackrel{\textstyle<}{\sim}$}}}

\begin{document}


\chapter{Pion Decay}\label{chap1}

\author[1,2]{Douglas Bryman}%
\author[3]{Robert Shrock}%


\address[1]{\orgname{University of British Columbia}, \orgdiv{Department of Physics and Astronomy}, \orgaddress{6224 Agricultural Road, Vancouver, BC V6T 1Z1 Canada}}
\address[2]{\orgname{TRIUMF}, \orgdiv{Physical Sciences Division}, \orgaddress{4004 Wesbrook Mall, Vancouver, BC V6T 2A3, Canada}}
\address[3]{\orgname{Stony Brook University}, \orgdiv{C. N. Yang Institute for Theoretical Physics and Department of Physics and Astronomy}, \orgaddress{Stony Brook, NY 11794 USA}}


\maketitle

\begin{abstract}[Abstract]
	 The decays of pions, the lightest particles composed of quarks, provide  important insight into fundamental questions in particle physics including  basic symmetries, the universality of fermionic weak interactions, and aspects of the strong interaction described by Quantum Chromodynamics (QCD).
  An 
  introduction to the theoretical and experimental study of charged and neutral pion decays is presented.
\end{abstract}

\begin{keywords}
 	pion\sep decay\sep  experiment\sep theory 
\end{keywords}

\begin{glossary}[Nomenclature]
	\begin{tabular}{@{}lp{34pc}@{}}
CVC& Conserved vector current \\
  JLAB& Thomas Jefferson National Accelerator Facility\\
  Hodoscope & A segmented instrument for observing the paths of subatomic particles\\
  LFU& Lepton flavor universality \\		
  LHC & Large Hadron Collider\\
  MWPC& Multi-wire proportional chamber\\
  PDG  & Particle Data Group \\
  PSI & Paul Scherrer Institute\\
  QCD& Quantum Chromodynamics\\
  TRIUMF & Canada's particle accelerator centre\\
	\end{tabular}
\end{glossary}

\section*{Objectives}
\begin{itemize}
	\item Illustrate how pion decays provide an abundant source of information on many aspects of elementary particle physics.
	\item Describe precise theoretical approaches  elucidating how the fundamentals of  pion decay have important connections to the weak, electromagnetic, and strong interactions.
	\item Show state-of-the-art experimental techniques that have been used to perform high-precision measurements.
	
\end{itemize}

\section{Introduction}\label{intro}

The pions are pseudoscalar mesons with spin-parity $J^P=0^-$ and valence quark content
\beq
|\pi^+\rangle: \ u\bar d \ , \quad |\pi^-\rangle: \ d\bar u
\label{pipm}
\eeq
and
\beq
|\pi^0\rangle: \ \frac{u\bar u - d \bar d}{\sqrt{2}} \ ,
\label{piz}
\eeq
where $u$ and $d$ are the quarks of the first generation with respective electric charges 2/3 and $-1/3$ (in units of $e$).
Pions play a very important role in nuclear and particle physics since pion exchange is a crucial part in the
nucleon-nucleon interaction that is responsible for the binding of
protons and neutrons in nuclei. This is indicated by the range of  the nucleon-nucleon force 
\beq
r_{\rm nuc} \simeq \frac{\hbar}{m_\pi c} \equiv \frac{1}{m_\pi},
\label{rnuc}
\eeq
where we use units with $c=\hbar=1$.  Indeed, in 1935, before
the discovery of the pions, Yukawa suggested that there should be a
strongly interacting particle with mass $m \simeq 1/r_{\rm nuc} \simeq          
10^2$ MeV whose exchange between nucleons would explain the range of
the nuclear force \cite{Yukawa:1935xg}. Shortly after Yukawa's
suggestion, Kemmer pointed out that the observed equality of the
strong force in proton-proton  and neutron-neutron
interactions implied that there should
be both charged and neutral strongly interacting mesons of
approximately the same
mass  \cite{Kemmer:1938sny}.  Charged pions were first
observed in 1947 by Powell and collaborators from analyses of cosmic
ray interactions in emulsions \cite{Lattes:1947mw,Lattes:1947mx}.  The neutral
pion was observed in 1950 in a photoproduction experiment by
Steinberger, Panofsky, and Steller \cite{Steinberger:1950equ,Panofsky:1950he}.

The experimental values of the charged and neutral pion masses given by the Particle Data Group
(PDG)~\cite{ParticleDataGroup:2024cfk} are
\beq
m_{\pi^\pm} = 139.57039 \pm 0.00017 \ {\rm MeV}
\label{mpip}
\eeq
and
\beq
m_{\pi^0} = 134.9768 \pm 0.0005 \ {\rm MeV}.
\label{mpiz}
\eeq
From conservation of the product of Charge, Parity, and Time symmetries (CPT), it follows that the mass $m_x$ and mean lifetime $\tau_x = \hbar/\Gamma_x$
where $\Gamma_x$  is the total decay rate,
of a particle $x$ and
its antiparticle, $\bar x$, are the same.
See Appendix ~\ref{Appendix} for additional theoretical information concerning pions and their interactions.
\footnote{For further background, the reader may wish to consult some modern textbooks on particle physics, including~  \cite{Donoghue:2022wrw,Langacker:2010zza,Quigg:2013ufa,Yndurain:2006amm,Pondrom:2023ekq}
and on quantum field theory, including
\cite{Itzykson:1980rh, Peskin:1995ev,Schwartz:2014sze,Weinberg:1996kr}.}

As  discussed  below,  $\pi^\pm$ decay weakly while
the $\pi^0$ decays electromagnetically.  The experimental values~\cite{ParticleDataGroup:2024cfk}
for their mean lifetimes are
\beq
\tau_{\pi^\pm} = 2.6033 \pm 0.0005) \times 10^{-8} \ {\rm s}
\label{taupip}
\eeq
and
\beq
\tau_{\pi^0} = 8.43 \pm 0.13 \times 10^{-17} \ {\rm s}.
\label{taupiz}
\eeq
%

The branching ratios of observed decay modes of
the $\pi^+$ are listed in Table ~\ref{pip_br_table}~\cite{ParticleDataGroup:2024cfk}. 
\begin{table}[htbp]
  \caption{ Branching ratios (BR) for observed pion decays
~\cite{ParticleDataGroup:2024cfk}.}
\begin{center}
\begin{tabular}{|c|c|} \hline\hline
  Mode             & BR  \\ \hline
$\mu^+ \nu_\mu$        & 99.98770(4) \ \%  \\
$\mu^+ \nu_\mu \gamma$ & $(2.00 \pm 0.25)  \times 10^{-4}$   \\
$e^+ \nu_e$            & $(1.23 \pm 0.004) \times 10^{-4}$   \\
$e^+ \nu_e \gamma$     & $(7.39 \pm 0.05)  \times 10^{-7}$   \\
$\pi^0 e^+ \nu_e$      & $(1.036 \pm 0.006)\times 10^{-8}$   \\
$e^+ \nu_e e^+e^-$     & $(3.2 \pm 0.5) \times 10^{-9}$      \\
\hline\hline
\end{tabular}
\end{center}
\label{pip_br_table}
\end{table}
Charged pions decay principally 
to muons ($\mu$) and muon neutrinos ($\nu_\mu$)\footnote{Muons ($m_\mu=105.6584$ MeV) and muon neutrinos ($\nu_\mu$) represent  the second generation of leptons which are fundamental SM color-singlet fermions. The third-generation leptons are the tau ($m_\tau=1776.9$ MeV) and the tau neutrino ($\nu_\tau$).} 
through the reactions 
  ($\pi_{\mu2}$)
\begin{equation}
\begin{aligned}
    \pi^+\rightarrow\mu^+ \nu_\mu \ {\rm and} \\
    \pi^-\rightarrow\mu^- \overline\nu_\mu  
\end{aligned}
    \label{pimu}
    \end{equation}
with the muon  and neutrino emitted with equal and opposite momentum $p=29.792$ MeV/c in the pion center of mass coordinate system.
The two-body electronic decay mode $\pi^+\rightarrow e^+ \nu_e$ ($\pi_{e2}$)  is suppressed to the level of $10^{-4}$ due to the vector (V) and axial-vector (A) left handed (V-A) nature of the weak interaction~\cite{Pondrom:2023ekq}. It was first observed in 1958~\cite{Fazzini:1958ii}\cite{Impeduglia:1958gha};shortly thereafter, its branching ratio relative to the $\pi^+ \to \mu^+ \nu_\mu$ decay mode,
\begin{equation}
 \remu
 =\frac{\Gamma(\pi^+\rightarrow e^+ \nu_e(\gamma))}{\Gamma(\pi^+\rightarrow \mu^+ \nu_\mu(\gamma))} \ ,
 \label{pienu(g)Br}
\end{equation} 
 was measured to a few percent accuracy~\cite{Anderson:1960zzb}~\cite{ DiCapua:1964zz, Bryman:1975yr} and subsequent experiments have reached 0.2\% precision~\cite{PiENu:2015seu}. The feebly interacting neutrinos involved in $\pi_{\mu2}$ and $\pi_{e2}$ decays
 are not observed in branching ratio measurements. The known mass eigenstates $\nu_i$, $i=1,2,3$ in the weak eigenstates $\nu_e$, $\nu_\mu$, and $\nu_\tau$ have sub-eV masses, which are negligibly small in pion decays. 
 Indeed, cosmological observations yield an upper bound 
 $\Sigma_{i=1}^3 m(\nu_i) < 0.12$ eV ~\cite{ParticleDataGroup:2024cfk}. The general expression for \remu~ in Eq. ~\ref{pienu(g)Br} includes radiative processes in which a gamma ray ($\gamma$) may  also be emitted and detected.

In the context of modern theoretical ideas, \remu~  is of interest as a test of the Standard Model (SM) hypothesis that all lepton flavors (i.e. e, $\mu$, and $\tau$) experience the same fundamental weak interaction strength, a principle known as lepton flavor universality (LFU). Rare radiative $\pi_{e2\gamma}$ decays are also  of interest for elucidating long-distance properties of the weak interactions i.e. those involving the quark structure or electromagnetic properties of the pion. Furthermore, the extremely rare pion beta decay, \pibeta~ ($\pi_{e3}$ ), 
is suppressed to the $10^{-8}$ level by phase space i.e. the small energy available for the decay, due the closeness of the  charged and neutral pion masses~\cite{ParticleDataGroup:2024cfk} 
\beq
\Delta_\pi=m_{\pi^+}-m_{\pi^0}=4.5936 \pm 0.0005 \ {\rm MeV} .
\label{Delta_mpi}
\eeq
%

 The observed decay modes of the  $\pi^0$ 
and its measured branching ratios are listed in Table \ref{piz_br_table}. 
\begin{table}[htbp]
  \caption{\footnotesize{ Branching ratios (BR) for observed $\pi^0$ decays
  ~\cite{ParticleDataGroup:2024cfk}, where the abbreviation
  Pos. stands for positronium, a bound state of $e^{+}e^{ - }$.}}
\begin{center}
\begin{tabular}{|c|c|} \hline\hline
Mode              & BR  \\ \hline
$\gamma\gamma$      & $(98.823 \pm 0.034)$ \%  \\
$e^+ e^- \gamma$    & $(1.174 \pm 0.035)$ \%   \\
$2(e^+ e^-)$        & $(3.34 \pm 0.16) \times 10^{-5}$ \\
$e^+ e^-$           & $(6.46 \pm 0.33) \times 10^{-8}$ \\
$\gamma+{\rm Pos}.$  & $(1.82 \pm 0.29) \times 10^{-9}$  \\
\hline\hline
\end{tabular}
\end{center}
\label{piz_br_table}
\end{table}
Neutral pions decay primarily to two photons via the reaction
\begin{equation}
\pi^0\rightarrow\gamma\gamma
\label{pi0}
\end{equation}
with 
 a small fraction  
 decaying via the three-body  Dalitz mode $\pi^0\rightarrow\gamma e^+ e^-$. Study of $\pi^0$ decay, in particular its lifetime, tests Quantum Chromodynamics (QCD) at confinement scale energies for  quarks bound within mesons.  A comprehensive review of the theory underlying neutral pion physics was presented in Ref. ~\cite{Bernstein:2011bx}. 

In the context of the SM, extraordinary precision, in some cases like \remu~ approaching $\pm 0.01\%$, has been obtained on predictions for  pion decay rates including radiative processes and higher-order effects. Experiments discussed below strive to reach similar levels of precision in order to confront the predictions to uncover any discrepancies which might imply the presence of new physics effects not included in the SM.

\section{Pion Decay Theory}


\subsection{Two-Body Leptonic Pion Decays}

\begin{figure}[htbp]
  \begin{center}
    \includegraphics[height=3cm,width=7cm]{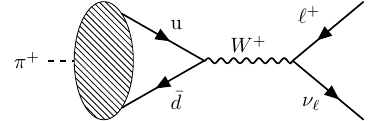}
  \end{center}
\caption{Feynman diagram for $\pi^+ \to \ell^+ \nu_\ell$, 
$\ell=\mu, \ e$.}
\label{pip_to_ellnu_figure}
\end{figure}

Since the calculations of $\pi^+ \to \mu^+ \nu_\mu$ and the
corresponding suppressed mode $\pi^+ \to e^+ \nu_e$ are the same, with
requisite changes for the outgoing lepton mass, we will
consider them together, labeling them as $\pi^+ \to \ell^+ \nu_\ell$,
where $\ell$ denotes $\mu$ or $e$\footnote{
In experiments on $\pi^+ \to \ell^+ \nu_\ell$ decays, the detectors usually collect 
events with  
photons emitted.  To indicate this, 
the decay rate with 
all
photons included is written as
$\Gamma(\pi^+ \to \ell^+ \nu_\ell (\gamma))$. Here we will take this as implicit in our discussion.  In Sec.~\ref{RadDecay} we will discuss radiative decays $\pi^+ \to \ell^+ \nu_\ell \gamma$, where the photon energy is above an experimentally determined cutoff (typically of order O(1) MeV). }.  In the Standard Model, these
decays occur at tree-level via the annihilation of the $u\bar d$ system in
the $\pi^+$ to yield a virtual $W^+$ vector boson that then materializes as
the $\ell^+ \nu_\ell$ pair in the final state, as shown in Fig. ~\ref{pip_to_ellnu_figure}. 

At the quark level, the
vertex for the annihilation $u \bar d$ of the (mass-eigenstates)
in the $\pi^+$ is described by the weak charged
current.  The charge-raising weak charged current is
\beqs
J_\lambda &=& \bar Q^{(up)}_{wk,L}  \gamma_\lambda Q^{(dn)}_{wk,L} 
= \bar Q^{(up)}_{m,L} V \gamma_\lambda Q^{(dn)}_{m,L} \ ,
\cr\cr
&&
\label{jwk}
\eeqs
where the subscripts (up) and (dn) denote charge 2/3 and charge $-1/3$
quarks and we have appended the subscripts $wk$ to indicate
weak-interaction (flavor) eigenstates and $m$ to indicate mass
eigenstates, with the flavor vectors $Q^{(up)}_m=(u,c,t)^T$ and $Q^{(dn)}_m=(d,s,b)^T$. In Eq. (\ref{jwk}), $\gamma_\lambda$ is a Dirac matrix (see Appendix ~\ref{Appendix}) and 
the unitary Cabibbo-Kobayashi-Maskawa (CKM)
quark mixing matrix
\beq
V = \left(\begin{array}{ccc}
  V_{ud} & V_{us} & V_{ub} \\
  V_{cd} & V_{cs} & V_{cb} \\
  V_{td} & V_{ts} & V_{tb} \end{array} \right ) 
\label{ckm}
\eeq
 encodes the effect of the unitary transformations
mapping the mass eigenstates of the charge $-1/3$ quarks and charge 2/3 quarks
to the respective weak-interaction 
eigenstates.
Hence, at the quark level, the Feynman vertex describing the annihilation
of the $u \bar d$ constituents of the $\pi^+$ is
\beq
\frac{g}{\sqrt{2}} \, V_{ud} \, [\bar v_{d,L}\gamma_\lambda u_{u,L}] \ ,
  \label{udvertex}
\eeq
where $g$ is the SU(2)$_L$ gauge coupling; $u_{u,L}$ and $\bar
v_{d,L}$ are the Dirac spinors for the $u$ and $\bar d$ quark and
antiquark, and the subscript $L$ indicates the left-handed chiral
components. From a combination of weak decay data, including, most
accurately, data on rates for superallowed nuclear beta decays, one finds ~\cite{ParticleDataGroup:2024cfk}\footnote{With
appropriate rephasings of quark fields, one can render $V_{ud}$ real
and positive, and we assume that this rephasing has been 
performed here. Also, note that the determination of $V_{ud}$ from the superallowed nuclear beta decays themselves yields $V_{ud}= 0.97367(32)$ ~\cite{ParticleDataGroup:2024cfk}.}
\beq
V_{ud} = 0.97435 \pm 0.00016.
\label{Vud}
\eeq

In the original Standard Model (SM),
the neutrinos were taken to be massless; in the  SM extended to
include nonzero neutrino masses,
given that the masses of the known
neutrino mass eigenstates are $\lsim O(1)$ eV,
we  neglect their masses.  In Sec.~\ref{Exotics}, the effects of possible additional neutrino mass eigenstates
with non-negligible masses will be discussed. In the SM, the leptonic
vertex is thus
\beq
\frac{g}{\sqrt{2}} \, [\bar u_{\nu_\ell,L}\gamma_\rho v_{\ell, L}] \ .
\label{leptonvertex}
\eeq
At the quark level, the amplitude is (dropping factors of $i$
that do not affect the final result)
\beq
{\cal M}(u \bar d \to \ell^+ \nu_\ell) = \Big ( \frac{g}{\sqrt{2}} \Big )^2 \,
V_{ud} \,
[\bar v_{d,L}\gamma_\lambda u_{u,L}] \Delta(q)^{\lambda\rho}
[\bar u_{\nu_\ell,L}\gamma_\rho v_{\ell,L}] \ ,
\label{qqamplitude}
\eeq
where $q$ is the four-momentum of the $\pi^+$ and
$\Delta^{\lambda\rho}(q)$ is the $W$ propagator.  In the unitary gauge,
the $W$ boson propagator ~\cite{Itzykson:1980rh} is

\beq
\Delta^{\lambda \rho}(q) =
\frac{-g^{\lambda\rho} + \frac{q^\lambda q^\rho}{m_W^2} }{q^2 - m_W^2} \ ,
\label{wpropagator}
\eeq
where $q^2$ is the Lorentz-invariant product
$q^2 \equiv q_\lambda q^\lambda = q_0^2 - |\vec q|^2$, with summation over repeated Lorentz indices
understood. Here, $q^2=m_{\pi^+}^2$.  In the propagator denominator, the $q^2$ term is negligibly small compared with the $m_W^2$ term, since $m_W = 80.4$ GeV and thus $q^2/m_W^2 =  m_\pi^2/m_W^2 = 3.0 \times
10^{-6}$.  Similarly, in the propagator
numerator, the $q^\lambda q^\rho/m_W^2$ contracts with the parent pion
momentum and the outgoing lepton momenta, yielding a term $m_\pi
m_\ell/m_W^2$, which is even smaller.  Hence, to high accuracy,
the $W$ propagator is just $\Delta^{\lambda\rho} = g^{\lambda
  \rho}/m_W^2$.  Factoring out the (1/2) factors from the chiral
projection operators (see Appendix ~\ref{Appendix}), one has
\beq
{\cal M}(u \bar d \to \ell^+ \nu_\ell) = \frac{g^2}{8m_W^2} \, V_{ud} \, 
[\bar v_{d}\gamma_\lambda(1-\gamma_5) u_{u}]
[\bar u_{\nu_\ell}\gamma^\lambda(1-\gamma_5) v_{\ell}] \ .
\label{qqamplitude2}
\eeq
In the SM (at tree-level as in Fig.~\ref{pip_to_ellnu_figure}), $g^2/(8m_W^2) = G_F/\sqrt{2}$, where
$G_F = 1.1663787(6) \times 10^{-5}$ GeV$^{-2}$ is the Fermi constant
(determined to highest precision from $\mu$ decay ~\cite{ParticleDataGroup:2024cfk}). 
Next, sandwiching the bilinear quark operator between the parent
$|\pi^+\rangle$ state
and the hadronic vacuum $\langle 0 |$, one has
\beq
\langle 0|[\bar v_{d}\gamma_\lambda(1-\gamma_5) u_{u}]|\pi^+\rangle =
if_\pi q_\lambda \ ,
\label{fpirel}
\eeq
which defines the pion decay constant $f_\pi$.\footnote{Note that some authors
write the right-hand side of Eq. (\ref{fpirel}) as
$i\sqrt{2}f_\pi q_\lambda$, so that the value of $f_\pi$ with their
normalization convention is 92 MeV.} Using the measured 
decay rate $\Gamma(\pi^+ \to \mu^+ \nu_\mu)$
~\cite{ParticleDataGroup:2024cfk},  $m_{\pi^+}$ and $m_\mu$,
and taking $G_F$ from $\mu$ decay and $|V_{ud}|$ from
superallowed nuclear beta decays, one can extract the value  $f_\pi = 130$ MeV. With the advent of highly
accurate lattice gauge theory simulations, the values of $f_\pi$ and 
other pseudoscalar decay constants have been 
calculated {\it ab initio} from the fundamental QCD Lagrangian, with
results in agreement with the older method, as reviewed, e.g., in
~\cite{ParticleDataGroup:2024cfk}. For example, a combination of recent results from
lattice groups yields the value ~\cite{FlavourLatticeAveragingGroupFLAG:2021npn}
\beq
f_\pi = 130.2 \pm 0.8 \ {\rm MeV}.
\label{fpi_lattice}
\eeq
The agreement of the experimental and ${\it ab~ initio}$ lattice QCD methods for calculating $f_\pi$ is an
important achievement. 
Combining these ingredients, one finally obtains the amplitude
\beq
{\cal M}(\pi^+ \to \ell^+ \nu_\ell) = \frac{G_F}{\sqrt{2}} \, V_{ud} \, f_\pi \,
[\bar u_{\nu_\ell} q \cdot \gamma (1-\gamma_5)v_{\nu_\ell}] \ .
\label{pidec_amplitude}
\eeq
Importantly, the $q \cdot \gamma = q^\mu\gamma_\mu$ term operating on the lepton spinor yields a
factor of $m_\ell$. Squaring the amplitude, taking the Dirac trace (see, e.g., ~\cite{Itzykson:1980rh}),
and performing the integration over final-state phase space (see the Appendix ~\ref{Appendix} for relevant general formulas), one obtains
the rate in the SM 
\beq
\Gamma(\pi^+ \to \ell^+ \nu_\ell) =
\frac{G_F^2 V_{ud}^2 f_\pi^2m_{\pi^+} m_\ell^2}{8\pi^2}
\bigg ( 1 - \frac{m_\ell^2}{m_{\pi^+}^2} \bigg )^2 \ .
\label{pipell2_decayrate}
\eeq
The appearance of the factor $m_\ell^2$ in the decay rate can be
understood on basic physical grounds. Since the $\pi^+$ has spin 0, by
angular momentum conservation, the final $\ell^+ \nu_\ell$ state must
also have angular momentum 0. Now (taking the neutrino masses to be
zero, so their helicities are the same as their chiralities), the $(V-A)$
charged current interaction produces a $\nu_\ell$ with helicity
$-1/2$.  Since the $\ell^+$ and $\nu_\ell$ move in opposite directions
in the rest frame of the parent $\pi^+$, this means that the $\ell^+$
must have the  wrong helicity, $-1/2$, whereas the $(V-A)$
interaction would naturally yield a helicity $+1/2$ for the $\ell^+$. In the decay process, the  wrong helicity is possible because of the nonzero lepton mass i.e. the factor $m_\ell$ in the amplitude. 
The degree of
helicity suppression can be seen equivalently by re-expressing the
decay rate in terms of the dimensionless variable
\beq
\delta^{\pi^+}_{\ell} = \frac{m_{\ell}^2}{m_{\pi^+}^2} \ ,
\label{delta_ell}
\eeq
namely
%
\beq
\Gamma(\pi^+ \to \ell^+ \nu_\ell) =
\frac{G_F^2 V_{ud}^2 f_\pi^2m_{\pi^+}^3}{8\pi^2} \, \delta^{\pi^+}_\ell \,
\bigg (1 - \delta^{\pi^+}_\ell \bigg )^2 .
\label{pipell2_decayrate2}
\eeq
%
For later reference, the total energy and magnitude of 3-momentum of the
$\ell^+$ in the rest frame of the parent pion are 
\beq
E_{\ell} = \frac{m_{\pi^+}^2+m_\ell^2}{2m_{\pi^+}}
\label{ell}
\eeq
and
\beq
|\vec p_\ell|=|\vec p_{\nu_\ell}| = \frac{m_{\pi^+}^2-m_{\ell}^2}{2m_{\pi^+}},
\label{pellnu}
\eeq
where we have dropped neutrino mass terms.

The ratio of the rates, or equivalently, the branching ratios, for the
decays $\pi^+ \to e^+ \nu_e$ and $\pi^+ \to \mu^+ \nu_\mu$ is
\beq
R_{e/\mu,SM} \equiv
\frac{\Gamma([\pi^+ \to e^+ \nu_e] + [\pi^+ \to e^+ \nu_e\gamma])}{\Gamma([\pi^+ \to \mu^+ \nu_\mu] + [\pi^+ \to \mu^+ \nu_\mu\gamma])} \equiv 
\frac{\Gamma(\pi^+ \to e^+ \nu_e (\gamma))}
{\Gamma(\pi^+ \to \mu^+ \nu_\mu (\gamma))} = 
\bigg ( \frac{m_e}{m_\mu} \bigg )^2  \frac{(1-\delta^{\pi^+}_e)^2}
  {(1-\delta^{\pi^+}_\mu)^2} \, \Big ( 1 + \delta_{RC} \Big ) \ ,
\label{Remu_SM}
\eeq
where $\delta_{RC}$ is the electromagnetic radiative correction.
Historically, early measurements of this ratio provided important evidence
in support of the $(V-A)$ structure of charged-current weak interactions
before the development of the full Standard Model (e.g. see ~\cite{Pais:1986nu} and
references therein).
Several  calculations of the radiative corrections have been performed 
~\cite{Berman:1958gx,Kinoshita:1959ha,Kinoshita:1958ru,Goldman:1976gh}.
These take into account emission of soft photons to match 
realistic experimental conditions. 
Using the most recent
 radiative correction calculations  ~\cite{Marciano:1993sh, Cirigliano:2007ga, Cirigliano:2007xi},\footnote{
  Hadronic structure-dependent bremsstrahlung (SDB) contributions to the
  radiative corrections are estimated to be negligibly small for
  $R^{(\pi)}_{e/\mu}$ and hence are not included in the calculations of
  ~\cite{Marciano:1993sh, Cirigliano:2007ga, Cirigliano:2007xi}.} 
the SM prediction for the ratio (\ref{Remu_SM}) is\footnote{Ref. ~\cite{Bryman:2011zz} estimated the uncertainty due
to the leading-log contributions to $\delta_{RC}$ differently from Refs.
~\cite{Cirigliano:2007ga, Cirigliano:2007xi,Bryman:2021teu}, leading to a SM-predicted ratio
$R_{e/\mu} = 1.23524(19) \times 10^{-4}$ with a larger estimated uncertainty.}
\beq
R_{e/\mu,{\rm SM}} = 1.23524(15) \times 10^{-4}
\label{Remu_SM_value}.
\eeq

\subsection{Radiative Decays $\pi^+ \to \ell^+ \nu_\ell \gamma$}
\label{RadDecay}

In this section we discuss radiative decays of charged pions, $\pi^+            
\to \ell^+ \nu_\ell \gamma$  ($\pi^+_{\ell 2 \gamma}$), where $\ell = \mu, \ e$.  
As noted above,
in an actual experiment on $\pi^+ \to \ell^+ \nu_\ell$ decays, the
detector collects not only events of this type, but also events
involving the emission of a photon of 
sufficiently low energy not to be observed.
Thus, when one
refers to decays of the form $\pi^+ \to \ell^+ \nu_\ell \gamma$, this
entails a criterion for the minimum energy of the
photon in a given experiment. 
In an experiment, one
can measure (a) the energy of the $\ell^+$, (b) the energy of the
photon, and (c) the angle $\theta_{\ell \gamma}$ between the 3-momenta
of the $\ell^+$ and the photon.

  The leading (tree-level) contributions to  $\pi^+ \to \ell^+ \nu_\ell\gamma$
  arise from three types of processes, in
  which the photon is radiated from (i) the initial $\pi^+$, (ii) the
  outgoing $\ell^+$, and (iii) intermediate hadronic states .
  Processes (i) and (ii) are called ``inner bremsstrahlung'' (IB)
  contributions, while (iii) is called the ``structure-dependent'' (SD)
  contribution. Referring to Fig. \ref{pip_to_ellnu_figure}, (i) corresponds to
  a $\gamma$ radiated from the dashed line representing the initial $\pi^+$,
  (ii) corresponds to a $\gamma$ radiated from the outgoing $\ell^+$,
  and (iii) corresponds to a $\gamma$ radiated from a process in which the $\gamma$ is radiated from the grey blob. In 
  the  SM, there would also be a contribution from a 
  process in which the $\gamma$
  is radiated from the virtual $W^+$ vector boson, but
this is negligibly small because of the suppression due to the resultant two $W$ propagators and is not included in the usual low-energy
effective field theory in which the $W$ has been integrated out to obtain the Fermi current-current Lagrangian.

Thus, the decay amplitude has the form
\beq
      {\cal M}(\pi^+ \to \ell^+ \nu_\ell \gamma) =
      {\cal M }_{IB} + {\cal M}_{SD} \ .
\label{amp_pi_ell_nu_gamma}
\eeq

The four-momenta of the parent
$\pi^+$ and the outgoing $\ell^+$, $\nu_\ell$, and $\gamma$ are denoted
$p_\pi$, $p_\ell$, $p_\nu$, and $k$, respectively, satisfying
\beq
p_\pi = p_\ell + p_\nu + k \ .
\label{pip_to_lng_momentum_con}
\eeq
We define the four-momentum transfer to the $\ell^+\nu_\ell$
pair as 
\beq
q = p_\ell + p_\nu  \ .
\label{qpp}
\eeq
The polarization four-vector for the outgoing photon is 
denoted $\epsilon_\nu^*(k)$. It is convenient
to introduce the two dimensionless Lorentz-invariant variables
\beq
x_\gamma = \frac{2p_\pi \cdot k}{m_{\pi^+}^2}
\label{xgamma}
\eeq
and
\beq
x_\ell = \frac{2p_\pi \cdot p_\ell}{m_{\pi^+}^2} \ .
\label{xell}
\eeq
We denote the $\ell^+$ and photon energies in the rest frame of the
decaying $\pi^+$ as $E_\ell$ and $E_\gamma$. In terms of these quantities,
\beq
x_\gamma = \frac{2E_\gamma}{m_{\pi^+}}
\label{xgamma_Egamma}
\eeq
and
\beq
x_\ell = \frac{2E_\ell}{m_{\pi^+}} \ .
\label{xell_Eell}
\eeq
The physical ranges of $x_\gamma$ and $x_\ell$ are
\beq
2\sqrt{\delta_\ell} \le x_\ell \le 1+\delta_\ell \ ,
\label{xell_range}
\eeq
(see Eq. (\ref{delta_ell})), and
\beq
1-\frac{x_\ell}{2}-\frac{1}{2}\sqrt{x_\ell^2-\delta_\ell} \le x_\gamma
\le
1-\frac{x_\ell}{2}+\frac{1}{2}\sqrt{x_\ell^2-\delta_\ell} \ .
\label{xgamma_range}
\eeq
Note that
\beq
q^2 = (1-x_\gamma)m_{\pi^+}^2 \ .
\label{qdq}
\eeq
In terms of $x_\ell$ and $x_\gamma$, the angle between ${\vec p}_\ell$
and ${\vec k}$ in the $\pi^+$ rest frame is given by
\beq
\cos\theta_{\ell\gamma} = \frac{x_\ell(x_\gamma-2)+2(1-x_\gamma+\delta_\ell)}
  {x_\gamma\sqrt{x_\ell^2-\delta_\ell}} \ .
\label{cos_theta}
\eeq

Normalizing by the total $\pi^+ \to \ell^+ \nu_\ell$ decay rate, the
resultant differential radiative decay rate has the form
\beq
\frac{1}{\Gamma(\pi^+ \to \ell^+ \nu_\ell)} \,
\frac{d\Gamma(\pi^+ \to \ell^+ \nu_\ell\gamma)}{dx_\gamma dx_\ell} =
a_{IB} f_{IB}(x_\gamma,x_\ell) +a_{SD}f_{SD}(x_\gamma,x_\ell)+
a_{int.}f_{int}(x_\gamma,x_\ell) \ ,
\label{piradform}
\eeq
where the three terms arise from $|{\cal M}_{IB}|^2$, $|{\cal M}_{SD}|^2$,
and the IB-SD interference (int.) term,
$2{\rm Re}({\cal M}_{IB}{\cal M}_{SD}^*)$, from Eq. (\ref{amp_pi_ell_nu_gamma}). 
The second and third terms in Eq. (\ref{piradform}) are negligibly small
in $\pi^+ \to \mu^+ \nu_\mu \gamma$ decay.
For details on the second two terms in Eq. (\ref{piradform}), see ~\cite{DeBaenst:1968wha}, ~\cite{Bryman:1982et}.
For both $\mu$- and $e$ type radiative decays, the coefficient $a_{IB}$ and
the function $f_{IB}(x_\gamma,x_\ell)$ in Eq. (\ref{piradform}) are given by ~\cite{Brown:1964zza,Neville:1961zz}
\beq
a_{SD} = \bigg ( \frac{\alpha_{em}}{2\pi}\bigg ) \, \frac{1}{(1-\delta_\ell)^2} \ ,
\label{aib}
\eeq
where $\alpha_{em} = e^2/(4\pi) = 1/137.036$ is the fine-structure constant, and
\beq
f_{IB}(x_\gamma,x_\ell) =
\bigg [ \frac{(1+\delta_\ell)-x_\ell}
  {x_\gamma^2[x_\gamma+x_\ell-(1+\delta_\ell)]} \bigg ]
\bigg [ x_\gamma^2+2(1-x_\gamma)(1-\delta_\ell) - \frac{2x_\gamma \delta_\ell (\
1-\delta_\ell)}{x_\gamma+x_\ell-(1+\delta_\ell)} \bigg ] \ .
\label{fib}
\eeq

Structure-dependent terms 
are valuable for the information that they yield on the hadronic aspects of the $u \bar d$ in the $\pi^+$ annihilation process and as inputs for chiral perturbation theory.  (For reviews of chiral perturbation theory, see  ~\cite{Weinberg:1996kr,Donoghue:1992dd,Pich:1995bw,Meissner:2024ona} and references therein.) The structure-dependent term in the decay amplitude depends on vector and axial-vector form factors, $F_V(q^2)$ and $F_A(q^2)$, and is given by
\beq
{\cal A}_{SD} = i\frac{e G_F V_{ud}}{\sqrt{2} m_{\pi^+}} \, [\bar u_\nu\gamma_\mu(1-\gamma_5)v_\ell] \, \epsilon^*_\nu(k) 
\bigg [ F_V(q^2)\epsilon^{\mu\nu\alpha\beta} 
p_{\pi\alpha}k_\beta + i F_A(q^2)\{ g^{\mu\nu}(p_\pi \cdot k) - k^\mu p_\pi^\nu \} \bigg ].
\label{m_sd}
\eeq
Let us define the ratio
\beq
r_F \equiv \frac{F_A(q^2)}{F_V(q^2)}.
\label{rf}
\eeq
Squaring and taking traces, one obtains the contribution to the differential decay rate from the SD term itself, as 
\beq
\frac{1}{\Gamma(\pi^+_{\ell 2})} \, 
\frac{d^2\Gamma(\pi^+_{\ell 2 \gamma})}{dx_\gamma dx_\ell}
= \bigg ( \frac{\alpha_{em}}{8\pi} \bigg )
\bigg ( \frac{m_{\pi^+}}{m_\ell} \bigg )^2 
\bigg ( \frac{F_V(q^2)}{f_\pi} \bigg )^2 
\bigg [ (1+r_F)^2T_+ + (1-r_F)^2 T_- \bigg ]
\label{d2gamma_sd}
\eeq
where
\beq
T_+ = (1-x_\gamma)(x_\gamma+x_\ell-1)^2 
\label{Tplus}
\eeq
and
\beq
T_- = (1-x_\gamma)(1-x_\ell)^2.
\label{T_minus}
\eeq
Because of the factor $(m_{\pi^+}/m_\ell)^2$ in the differential decay rate Eq. (\ref{d2gamma_sd}), the SD contribution is much larger in $\pi^+_{e 2\gamma}$ decay than in $\pi^+_{\mu 2 \gamma}$ decay.  
The conserved vector current (CVC) hypothesis\cite{Pais:1986nu} predicts 
$F_V(0)=0.0259(5)$ \cite{ParticleDataGroup:2024cfk}.  Chiral perturbation theory predictions for $F_A(0)$ are in the range 
0.01-0.02 \cite{Holstein:1986uj,Bijnens:1996wm,Geng:2003mt}.

\subsection{Pion Beta Decay $\pi^+ \to \pi^0 e^+ \nu_e$} 
\label{pibeta_theory}

The charged pion undergoes a tree-level weak charged-current decay
$\pi^+ \to \pi^0 e^+ \nu_e$, due to a quark level transition
$\bar d \to \bar u + e^+ \nu_e$. 
This has the very small branching ratio listed in Table \ref{pip_br_table}, owing to the small mass difference $\Delta_\pi$ (Eq.~\ref{Delta_mpi}). 
The decay occurs between two
spin-0 members of an (flavor SU(2)) isotriplet and hence involves only
the vector part of the charged weak current,
analogous
to superallowed nuclear beta decays~\cite{Hardy:2020qwl}. 
Defining the dimensionless variable
\beq
\epsilon \equiv \bigg (\frac{m_e}{\Delta_\pi} \bigg )^2,
\label{epsilone}
\eeq
the SM prediction for the rate is
~\cite{Kallen:1964lxa,Sirlin:1977sv,Sirlin:1981ie}
\beq
\Gamma_{\pi \beta, SM} = \frac{G_F^2 V_{ud}^2 \Delta_\pi^5}{30\pi^3} \,
\bigg ( 1 - \frac{\Delta_\pi}{2m_{\pi^+}} \bigg )^3f(\epsilon)(1+\delta_r) \ ,
\label{gamma_pibeta}
\eeq
where
\beq
f(\epsilon) = (1-\epsilon)^{1/2}
\bigg (  1 - \frac{9}{2}\epsilon - 4\epsilon^2 \bigg )
  + \frac{15}{2} \epsilon^2
  \ln \bigg ( \frac{1+\sqrt{1-\epsilon}}{\sqrt{\epsilon}} \bigg ) -
  \frac{3}{7}\bigg (\frac{\Delta_\pi}{m_{\pi^+} + m_{\pi^0}} \bigg )^2 
\label{fpib}
\eeq
and $\delta_r$ incorporates radiative corrections~\cite{Sirlin:1977sv,Cirigliano:2002ng,Czarnecki:2019iwz,Feng:2020zdc,Yoo:2023gln}. The calculations in Refs.~\cite{Cirigliano:2002ng,Czarnecki:2019iwz} both yielded the result $\delta_r=0.0334(10)$. A recent lattice calculation \cite{Feng:2020zdc} (in agreement with a subsequent lattice computation ~\cite{Yoo:2023gln}) has obtained $\delta_r = 0.0332(1)(3)_{\rm HO}$, where "HO" stands for an estimate of higher-order contributions. 
The largest contributions to the fractional uncertainty in the SM prediction of $\Gamma_{\pi \beta,SM}$ Eq.~\ref{gamma_pibeta} are approximately 
 $6 \times 10^{-4}$ from $\Delta_\pi^5$ and  $3.3 \times 10^{-4}$ from $|V_{ud}|^2$ resulting in an overall
 uncertainty of $0.07 \%$.
 Then, the SM prediction is obtained as 
$BR(\pi^+_{e3}) = \Gamma_{\pi\beta,SM}/\Gamma_{\pi^+} = 
\Gamma_{\pi\beta,SM} \, \tau_{\pi^+}$: 
%
\beq
BR(\pi^+ \to \pi^0 e^+ \nu_e,SM) = 
\bigg [ [(1.0390 \pm 0.0008) 
\times 10^{-8} \bigg ] \, \bigg ( \frac{V_{ud}}{0.97435}\bigg )^2.
\label{pibeta_br_sm}
\eeq
Here, the dependence on $|V_{ud}|$ is shown explicitly using  the central value  from a global CKM quark mixing matrix fit
\cite{ParticleDataGroup:2024cfk}. However,  there is currently tension among values extracted via different methods, as discussed in ~\cite{ParticleDataGroup:2024cfk}.

\subsection {Neutral Pion Decay }
\label{pi0_theory}

\begin{figure}[htbp]
  \begin{center}
    \includegraphics[height=3cm,width=7cm]{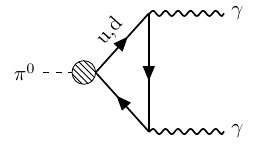}
  \end{center}
\caption{Feynman diagram for $\pi^0 \to \gamma\gamma$. For Bose symmetry, the corresponding diagram with the photon lines interchanged is also added.}
\label{piz_to_2gamma_figure}
\end{figure}

In contrast to the charged pions, which decay weakly, the $\pi^0$
decays electromagnetically, and hence its mean lifetime, $\tau_{\pi^0}$, given in Eq. (\ref{taupiz}), is much shorter than $\tau_{\pi^+}$. 
The amplitude for $\pi^0 \to \gamma\gamma$ arises from the 
triangle diagram shown in Fig. \ref{piz_to_2gamma_figure} and the corresponding diagram with photon lines interchanged, to maintain Bose symmetry of the final state. (This symmetry requires that the wavefunction for a set of identical bosons must be symmetric under interchange of the bosons.) 
In Fig. ~\ref{piz_to_2gamma_figure}, the dash-filled oval  (blob) represents the contribution from the anomalous
divergence in the axial-vector current
~\cite{Adler:1969gk,Bell:1969ts,Adler:1969er}. Defining the 
axial-vector current with global isospin index $a$ as 
$J_{5\mu}^a(x) = \bar\psi T^a \gamma_\mu\gamma_5 \psi$, where 
$T_a=(1/2)\tau^a$ ($\tau^a$ is a Pauli matrix, as listed in  Appendix ~\ref{Appendix}) is a generator
of isospin SU(2), this anomalous divergence is 
\beq
(\partial^\mu J_{5\mu}^3)_{\rm anom.} = \frac{\alpha_{em}}{4\pi} F_{\alpha\beta}\tilde F^{\alpha\beta} \ ,
\label{anomdiv}
\eeq
where $\tilde F^{\alpha\beta} = (1/2)\epsilon^{\alpha\beta\mu\nu} F_{\mu\nu}$.  
$\epsilon^{\alpha\beta\mu\nu}$ is the totally antisymmetric tensor density defined such that $\epsilon^{0123}=1$, 
and $F_{\mu\nu}$ is the 
electromagnetic field strength tensor (see, e.g. ~\cite{Itzykson:1980rh,Donoghue:2022wrw,Bernstein:2011bx}). This amplitude is
\beq
 {\cal M}(\pi^0 \to \gamma\gamma) = \frac{\alpha_{em} N_c}
 {8\pi \sqrt{2} f_\pi}(q_u^2 - q_d^2) \epsilon^{\mu\nu\alpha\beta} F_{\mu\nu}F_{\alpha\beta} \ ,
\label{amp_piz_to_2gamma}
\eeq
where $q_u$ and $q_d$ are the electric charges of the $u$ and $d$ quarks in units of $e$, and
$N_c$ is the number of quark colors. Substituting the values
$N_c=3$, $q_u=2/3$, and $q_d=-1/3$ and doing the integral over
phase space (with a $1/2$ factor to take account of identical particles in the final state), this yields the rate
\beq
\Gamma(\pi^0 \to \gamma\gamma) = \frac{\alpha_{em}^2 m_{\pi^0}^3}{32\pi^3 f_\pi^2}.
\label{gamma_piz_to_2gamma}
\eeq
Although the axial anomaly is not renormalized by higher-order
corrections ~\cite{Adler:1969er},
there are corrections to this decay rate owing to small violations of chiral
symmetry and isospin symmetry. These are estimated to increase the calculated decay
rate by approximately 4.5 \% ~\cite{Bernstein:2011bx} to
$\Gamma(\pi^0 \to \gamma\gamma) = 7.74 \pm 0.37$ eV, corresponding to
$\tau_{\pi^0} = 0.80 \times 10^{-16}$ s.
Historically, the measured value of $\Gamma(\pi^0 \to \gamma\gamma)$ decay was compared with the prediction based on the amplitude
(\ref{amp_piz_to_2gamma}) with $q_u=2/3$ and $q_d=-1/3$ but with $N_c$ as a free parameter to infer evidence for the value $N_c=3$. 
However, this was actually not consistent. Cancellation of chiral triangle anomalies ~\cite{Itzykson:1980rh,Pais:1986nu} in the Standard Model implies that 
$N_cY_{Q_L} + Y_{L_L}=0$, where $Q_L={u \choose d}_L$ and $L_L ={\nu_e \choose e}_L$ are the SU(2)$_L$ doublets for the quarks
and leptons 
and $Y_{Q_L}$ and $Y_{L_L}$ are the weak hypercharges of these
doublets satisfying $q_{em} = T_{3L} + (Y/2)$. Here, a SM 
fermion transforms under the weak isospin SU(2)$_L$ factor 
group in the SM gauge group as $(T_L,T_{3L})$. 
Then,  $q_u = q_d+1$ and $q_\nu = q_\ell + 1$.  With $q_\nu=0$, this implies that ~\cite{Abbas:1990kd,Gerard:1995bv,Chow:1995by,Shrock:1995bp,Bernstein:2011bx}
\beq
q_u = \frac{1}{2}\bigg ( 1 + \frac{1}{N_c} \bigg ) \ ,  \quad\quad  q_d = \frac{1}{2}\bigg ( -1 + \frac{1}{N_c} \bigg ) \ .
\label{anomaly_quqd}
\eeq
Now from (\ref{anomaly_quqd}) it follows that 
$q_u^2 - q_d^2 = 1/N_c$.
Therefore, Eq.~\ref{amp_piz_to_2gamma} is independent of $N_c$.
Note that the baryon spectroscopy argument for $N_c=3$ is sufficient by itself, as discussed in  Appendix ~\ref{Appendix}.

\section{Pion Decay Experiments}\label{Exp}

In this section a brief review of the most recent experiments on pion decays is presented.

\subsection{Charged Pion Decay Experiments}\label{ExpCharged}

Experiments dealing with the weak interaction properties of charged pions have mostly employed accelerator-produced low-energy beams of positive pions  (typically, with kinetic energy $T_{\pi^+} \approx 30$ MeV) which were degraded by passing through absorbers and  stopped in an experimental target where their decay products were observed with nearby detectors.\footnote{Although the same results for the decays of negative pions would be expected, $\pi^-$ particles stopped in matter form pionic atoms and quickly undergo nuclear capture~\cite{Itahashi:2023boi} making them unsuitable for such studies.}
Subsequent generations of experiments have continued to improve precision by increasing statistics through the use of more intense beams and detectors with larger efficiencies or acceptances, coupled with the evolution in  understanding and control of dominant systematic uncertainties studied with larger data sets and improved simulations.

\subsubsection{$\pi^+ \rightarrow e^+ \nu_e$}\label{PIENU}

\begin{figure}[H]
	\centering
\includegraphics[width=0.7\textwidth]{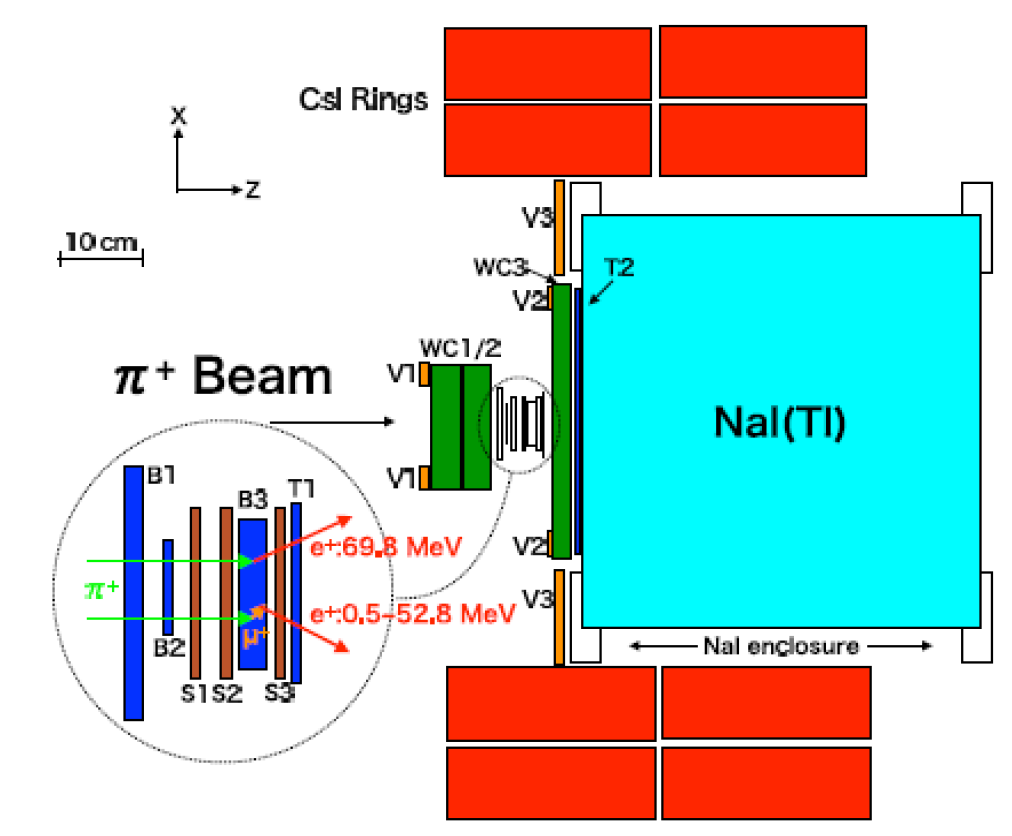}
	\caption{The experimental setup for the TRIUMF PIENU~\cite{PiENu:2015seu} experiment. The
cylindrical NaI(Tl) crystal is surrounded by a cylindrical array of
CsI crystals~\cite{Ito_Priv_Comm}.
}
	\label{pienu_setup}
\end{figure}

The PIENU~\cite{PiENu:2015seu} experiment at TRIUMF  produced the most recent measurement of  \remu. The experimental setup is shown in  Fig.~\ref{pienu_setup}.  The branching ratio \remu~ is obtained from the ratio of
positron yields from \pienu~ decays (mono-energetic positron with total
energy $E_e=69.8$ MeV) and  \pimu~ 
decays
followed by \mudecay~ decay with a mean lifetime of 2.2 $\mu$s. The \pimue~ decay chain results in positrons with  energy ranging from 
$E_e=0.5$ to $52.8$ MeV.  By measuring positrons and associated photons for each element of the ratio \remu~ with no magnetic field present, potential systematic uncertainties such as the numbers and location of pion stops, geometric acceptance for decay positrons, detection efficiency, and timing uncertainties for $\pi_{e2}$ and \pimue~ events mainly cancel, despite the large range of energies from 0.5 to 70 MeV. Small corrections, typically $<1\%$, are required to address energy-dependent cross sections for processes like multiple Coulomb scattering, Bhabha scattering 
($e^+ + e^- \rightarrow e^+ + e^-$), 
and positron annihilation in flight on atomic targets 
($e^+ + e^- \rightarrow \gamma$). 
Using  calorimeters with thickness of many radiation lengths ($X_0$)\footnote{The thickness of calorimeters measuring electromagnetic radiation like electrons, positrons, and photons
is measured in radiation lengths ($X_0$), which is the mean length in the material at which the energy of the
incoming radiation is reduced by a factor $1/e$ .} to measure energy ensures near uniform response for electrons, positrons, and photons over a wide energy range.
This technique for measuring \remu, pioneered by Di Capua et al.~\cite{DiCapua:1964zz}, has been used by all subsequent \pienu~ experiments~\cite{ParticleDataGroup:2024cfk}.

 A $\pi^+$ beam~\cite{Aguilar-Arevalo:2009ogf} at TRIUMF (momentum $p=75$ MeV/c, kinetic energy $T=19$ MeV) was  incident on the PIENU detector~\cite{PiENu:2015pkq} (Fig.~\ref{pienu_setup}),
degraded by two thin plastic scintillators B1 and B2, and
stopped in an 8-mm-thick scintillator target (B3) at a rate of
$5 \times 10^4 $ Hz. Pion tracking was provided by wire chambers
(WC1 and WC2) at the exit of the beam line and two
(x,y) sets of single-sided 0.3-mm-thick planes of Si-strip detectors S1 and S2 located immediately upstream
of B3. The positron calorimeter, 19 $X_0$ thick,
placed on the beam axis consisted of a 48-cm (diam.)
×~48-cm (length) single-crystal NaI(Tl) detector 
preceded by two thin plastic scintillators (T1 and T2).
Two concentric layers of pure CsI crystals  (9 $X_0$
radially) were used 
to capture electromagnetic showers due to positron interactions in the NaI(Tl) detector. Positron tracking was
done by an (x;y) pair of Si-strip detectors (S3) and multiwire proportional 
chambers (MWPC, WC3) in front of the  crystals.
A positron signal defined by a T1 and T2 coincidence,
occurring in a time window 
$t=- 300$ to 
$t=540$ ns with respect to
the incoming pion at $t=0$, was the basis of the main trigger logic.
Events in an early time window from
$t=6$ to 36 ns and high-energy (HE) events  with $Ecut > 52$ MeV 
in the calorimeter provided other triggers (early and HE), which included most \pienu~ decays. 

Events originating from stopped pions were selected
based on their energy losses in B1 and B2. Any events with
extra activity in the beam and positron counters (B1, B2,
T1, and T2) in the time region of $- 7$ to 1.5 $\mu$s with respect
to the pion stop were rejected. About 40\% of events
survived the cuts. A spatial fiducial cut for positrons entering
the NaI(Tl) detector required a track at WC3 to be within
60 mm of the beam axis to reduce electromagnetic shower
leakage from the crystals.

\begin{figure}[H]
	\centering
\includegraphics[width=1.\textwidth]{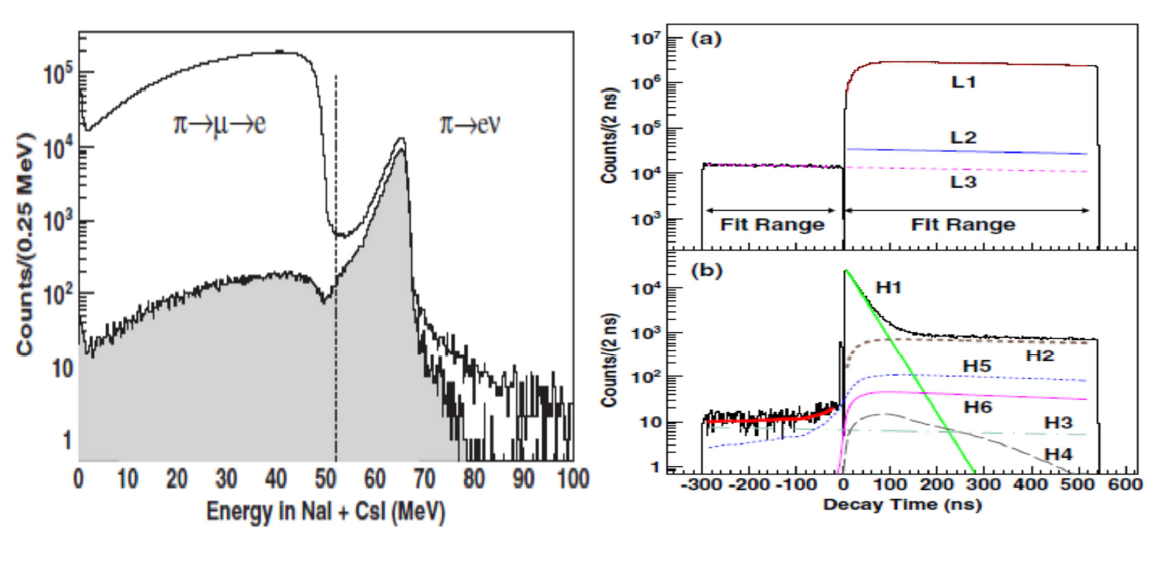}
	 \caption{ (Left) Energy spectra of positrons in the time region 5–35 ns
without and with (shaded) background-suppression cuts (see 
text). The vertical line at 52 MeV indicates the $E_{\rm cut}$ position. (Right)
Time spectra of positrons (thin line
histograms) in the (a) low- and (b) high-energy regions separated
at Ecut. 
In the low-energy time spectrum, the main components
were  \pimue~ decays at rest (L1), \mudecay~
decays (L2, after decays in flight of
pions ($\pi$DIF) and decays coming from previously stopped
(“old”) muons remaining in the target area (L3).  The primary time distribution component in the
high-energy region was the \pienu~ decay. The thick solid line in
(b) for $t < 0$ ns shows the fit. The fit for the other regions is
almost indistinguishable from the data and is omitted here. 
From  Ref.~\cite{PiENu:2015seu}. }

	\label{pienu_data}
\end{figure}

The summed NaI(Tl) and CsI energy for positrons in
the time region 5 to 35 ns is shown in Fig. \ref{pienu_data} (left). 
The time spectra
for events in the low- and high-energy regions separated at
$E_{\rm cut} =52$ MeV are shown in Fig. \ref{pienu_data} (right). 
Using a data set with
$4\times 10^5$ \pienu~ events after selection criteria ("cuts") were applied, a raw branching ratio was
determined from the simultaneous fit of the timing
distributions in Fig.~\ref{pienu_data}. To reduce possible bias, the raw branching
ratio was shifted (“blinded”) by a hidden random value
within 1\%. Prior to unblinding, all cuts and corrections
were determined, and the stability of the result against
variations of each cut was reflected in the systematic
uncertainty estimate.

Some corrections applied to the raw branching ratio
relied on simulations e.g. small energy-dependent effects in the
energy-loss processes of positrons change the relative
acceptances of low- and high-energy events, and the ratio of
the acceptances of the \pimue~ events and \pienu~
decays was found to be $0.9991 \pm 0.0003_{syst}$. 
The largest correction to the raw branching ratio was for
the \pienu~ events below $E_{\rm cut}$ which primarily arose
from the response function of the calorimeter i.e. the incomplete containment of electromagnetic showers.
The response function, complicated by the presence of  photonuclear
interactions, was measured separately~\cite{Aguilar-Arevalo:2010mkn}. The fraction of the
\pienu~ events below $E_{\rm cut}$ was estimated to be 
$3.16\pm0.12$
using simulations and  the \pimue~ background-suppressed energy spectrum shown as the shaded region  in Fig. ~\ref{pienu_data}~\cite{PiENu:2015seu}; \pimu~ background suppression was achieved  using an early
decay-time region 5 to 35 ns, pulse shape and total energy measurements in
B3 to identify incident muons,  and  beam particle  tracking measurements to identify \pimu~ decays in flight before stopping\cite{PIENU:2011aa}. 
After unblinding, the measured
branching ratio is 
$R^{\rm PIENU}_{e/\mu} =(1.2344 \pm 0.0023{stat}\pm
0.0019{syst})\times 10^{-4}$~\cite{PiENu:2015seu}. 

The PDG~\cite{ParticleDataGroup:2024cfk} average including  previous experiments done at TRIUMF~\cite{PiENu:2015seu,Bryman:1985bv,Britton:1993cj} and the Paul Scherrer Institut (PSI)~\cite{Czapek:1993kc} is
\begin{equation}
\label{Remu_exp_pdg}
     R^{\rm\overline{Exp}}_{e/\mu}= (1.2327 \pm 0.0023) \times 10^{-4}.
\end{equation}
Comparison between theory and experiment\cite{Bryman:2021teu} given in 
Eq.~(\ref{Remu_SM}) and Eq.~(\ref{Remu_exp_pdg}) provides 
the most
stringent test of the 
SM assumption for 
$e$--$\mu$ universality of the weak interaction. The results can be expressed in terms of the effective couplings $A_\ell$ ($\ell= e, \mu,\tau$) multiplying the low-energy charged current  contact interaction 
\begin{equation}
   L_{\rm CC}= A_\ell [\bar u \gamma^\mu P_L d]
                      [\bar \nu_\ell \gamma_\mu P_L \ell] \,,  
\end{equation}
where $P_L \equiv (1-\gamma_5)/2$. 
In the SM at tree level the couplings are given by  $A_\ell$ 
and satisfy LFU, i.e. $A_\ell / A_{\ell^\prime} = 1$.  
The measurement of $R^\pi_{e/\mu}$ results in 
 \begin{equation}
     \left( \frac{A_\mu}{A_e}  \right)_{ R^{\pi}_{e/ \mu} }= 1.0010 \pm 0.0009  \,
     \label{gpi}
 \end{equation}
which is in excellent agreement with the SM expectation. 
A deviation from $A_\ell / A_{\ell^\prime} = 1$ can originate from various mechanisms. In the literature it is common to interpret deviations in terms of flavor-dependent couplings $g_\ell$ of the $W$-boson to 
the leptonic current, in which case $A_\ell \propto g_\ell$.

\subsubsection{Radiative Pion Decay: ~$\pi^+\rightarrow e^+ \nu \gamma$ }\label{PiBeta_1}

\begin{figure}[H]
	\centering
\includegraphics[width=0.7\textwidth]{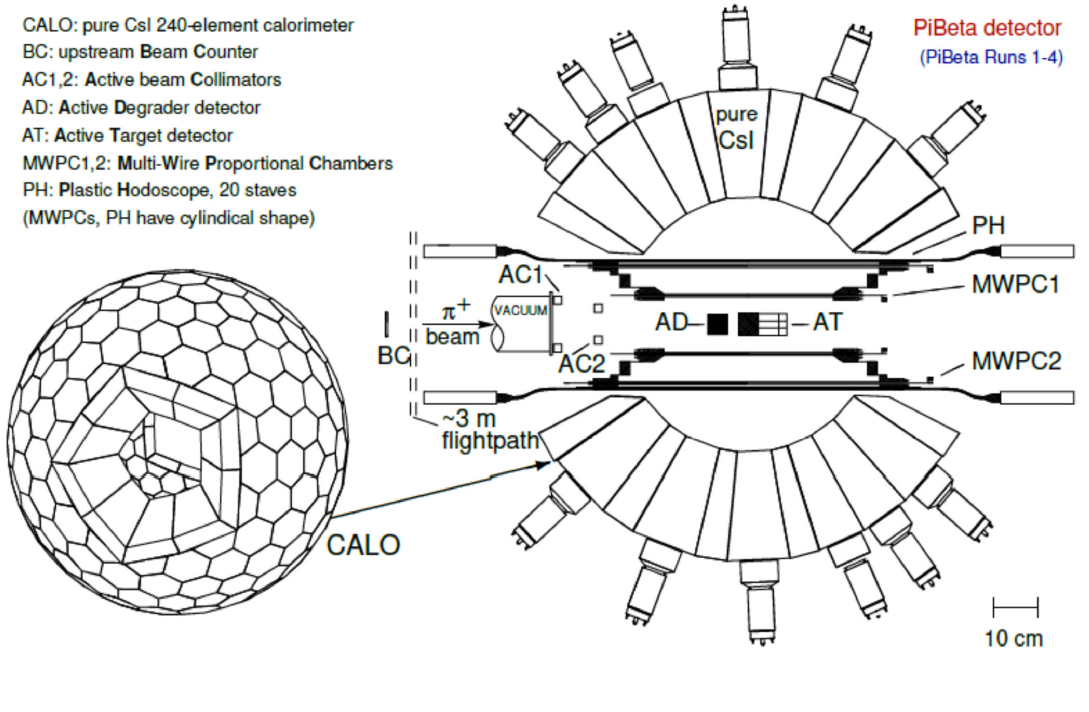}
	 \caption{Illustration of the PiBeta apparatus (see  text). Figure from ~\cite{Pocanic:2021zhy}.}
	\label{PiBetaAp}
\end{figure}

\begin{figure}[H]
	\centering
\includegraphics[width=0.5\textwidth]{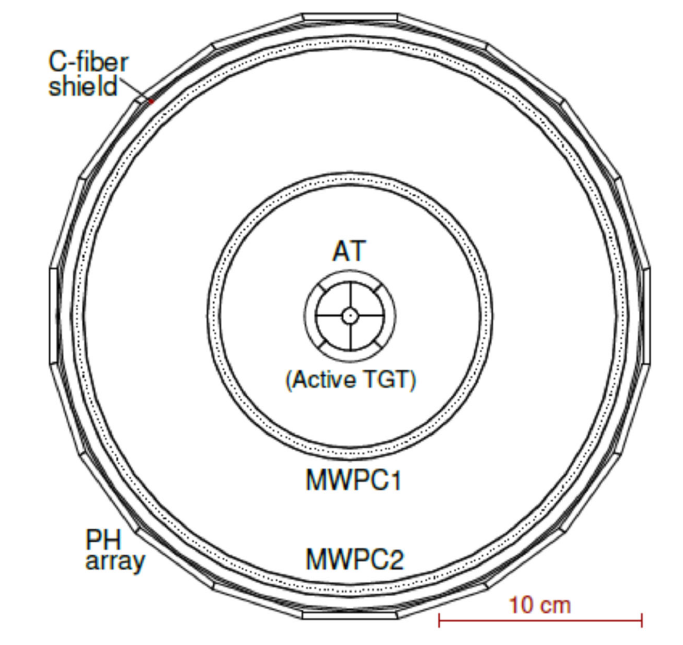}
	 \caption{Axial (beam) view of the
central detector region. 
Outward from center: (i) 
the 9-element segmented active target AT, (ii) cylindrical MWPC1 and MWPC2 trackers, (iii) thin 
cylindrical carbon-fiber shield around MWPC2, and (iv) the 20-element plastic scintillator hodoscope (PH) 
array.  AT outer ring elements were used for decay particle tracking. The BC, AD, AT and PH 
detectors were made of  plastic scintillator.
 Figure from ~\cite{Pocanic:2021zhy}.
}
	\label{PiBeta_end}
\end{figure}

The most recent experiment studying radiative pion decay ($\pi_{e\gamma}$) $\pi\rightarrow e \nu \gamma$ \cite{Bychkov:2008ws, Pocanic:2014jka} and pion beta decay ($\pi_{e3}$) $\pi^+\rightarrow \pi^0 e^+ \nu_e$ 
~\cite{Pocanic:2003pf} was the PiBeta experiment performed at  PSI~\cite{Frlez:2003vg}.
The PiBeta apparatus, schematically shown in Figs. ~\ref{PiBetaAp} and ~\ref{PiBeta_end}, detected $\pi^+$ decays from pions at rest in an
 active target (AT) at the center of a pure CsI 
  calorimeter with  240 crystal modules arranged  in a spherical configuration.\footnote{A slightly different configuration was used for part of the low-rate running~\cite{Frlez:2003vg}.} 

Referring to Fig. ~\ref{PiBetaAp}, incident beam pions from the $\pi$E1 channel at PSI with momentum $114$ MeV/c were detected by 
scintillation detectors (BC and AD, separated by 3.5m) and stopped in a segmented target (AT),  which 
was surrounded by two MWPC tracking detectors, and a  20-element plastic scintillator hodoscope. The apparatus acquired data during a “$\pi^+$-stop” gate spanning
-50 to 200 ns relative to the  pion stop time; a region of 10 ns from the pion stop time was excluded to avoid prompt hadronic pion interactions.
A crystal module  hit centrally by a 70 MeV positron would contain  about 90\% of the  energy. 
The location and energy of each
electromagnetic shower produced by a positron or photon in the calorimeter were extracted for trigger purposes from analog signal
sums of overlapping clusters of 7 to 9 modules. Twelve trigger configurations, combining
calorimeter and beam detector hit patterns of interest, were used to study signal and  background processes~\cite{Bychkov:2008ws}.

\begin{figure}[H]
	\centering
\includegraphics[width=0.5\textwidth]{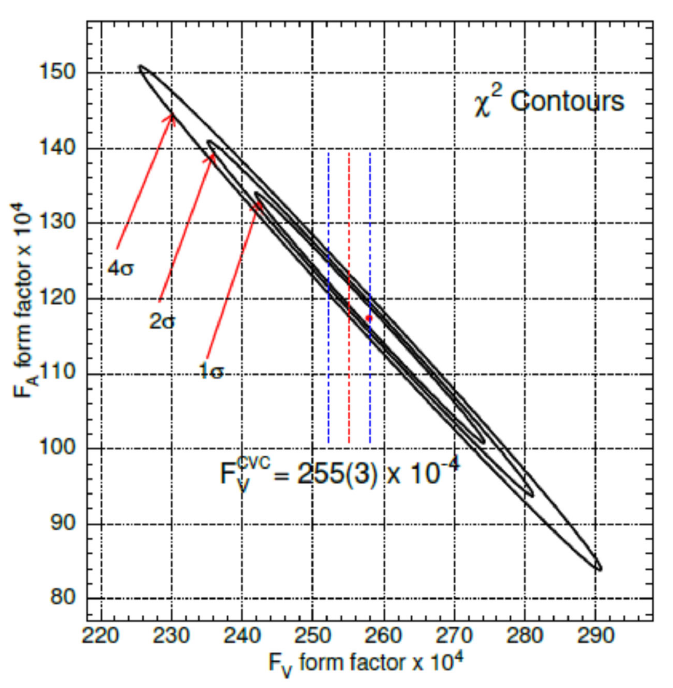}
	 \caption{$F_A$ and $F_V$ contours
for the  PiBeta $R_{e\gamma}$
 data set. Figure from ~\cite{Pocanic:2021zhy}.}
	\label{FAFV}
\end{figure}

PiBeta measurements were carried out in four run periods  including three at high stop rate 
with  $10^6$/s 
and  one  with $10^5$/s focused
on  $\pi_{e2\gamma}$ decays. 
The low-rate run allowed the 
precise
calibration and study of subtle calorimeter gain differences in  low- and high-threshold triggers. Analysis results for $F_A$ vs $F_V$ (see Sec.~\ref{RadDecay}) using  the full  data set
  are shown in Fig.~\ref{FAFV}.
  The compressed shape of
the  ellipse reflects the O(1\%) precision of the measurement of $F_A + F_V$ (SD$^+$),
and the  lower sensitivity to $F_A - F_V$ (SD$^-$).

The measured branching ratio depends on the ranges in $E_\gamma$, $E_e$ and (the related) range in the opening angle $\theta_{e\gamma}$ between ${\vec p}_e$ and ${\vec p}_\gamma$. The experiment ~\cite{Bychkov:2008ws} used several ranges and derived resultant values of $BR(\pi^+_{e2\gamma})$, all of which were in agreement with theoretical predictions. The value listed in Table \ref{pip_br_table}, refers to the range $E_\gamma > 10$ MeV and $\theta_{e\gamma} > 40^\circ$. 
Concerning form factors discussed in Sec.~\ref{RadDecay}, a fit was performed with the form
\beq
F_V(q^2) = F_V(0)\bigg (1+a\frac{q^2}{m_{\pi^+}^2} \bigg ), \quad F_A(q^2)=F_A(0)
\label{fvafit}
\eeq%
where $a$ is a slope parameter describing the Taylor series
expansion of the vector form factor as a function of 
$q^2/m_{\pi^+}^2$,
and obtained the results 

\beq
F_V(0) = 0.0258 \pm 0.0017, \quad F_A(0) = 0.0119 \pm 0.0001
\label{fvfa_exp}
\eeq
with 
\beq
a = 0.10 \pm 0.06.
\label{avalue}
\eeq

There is qualitative agreement of the experimental result with theoretical expectations~\cite{Mateu:2007tr} for $F_V(0)$ and $F_A(0)$ indicated in Sec.~\ref{RadDecay}.
\footnote{Fig.~\ref{FAFV} shows an earlier estimate for $F_V(0)$.}  
Analysis of the integral 
 decay rate yielded the branching ratio for the
kinematic region $E_\gamma
 > 10 $ MeV and $\theta_{e-\gamma}
 > 40^\circ$:  $B^{exp}_{R_{e\gamma}} = 73.86(54) \times 10^{-8}$ ( O(1\%) precision). The differential decay rate also   led to 
a stringent limit on a possible admixture of a tensor interaction~\cite{Bhattacharya:2011qm}  $-5.2\times 10^{-4} < F_T < 4.0 \times 10^{-4}$
at 90\% confidence level.

The branching ratio for $\pi^+ \to \mu^+\nu_\mu\gamma$ was measured most recently in an experiment at PSI ~\cite{Bressi:1997gs}. This experiment measured both $E_\mu$ and $E_\gamma$, and in the latter case, reached down to $E_\gamma=1$ MeV.  The result, $BR(\pi^+_{\mu e \gamma}; E_\gamma > 1 \ {\rm MeV}) = (2.0 \pm 0.24 \pm 0.08) \times 10^{-4}$, was in agreement with the SM prediction 
$BR^{SM}(\pi^+_{\mu e \gamma}; E_\gamma > 1 \ {\rm MeV}) =2.283 \times 10^{-4}$~\cite{Bressi:1997gs}.

\subsubsection{Pion Beta Decay}

\label{PiBeta_2}
\begin{figure}[H]
	\centering
\includegraphics[width=1.\textwidth]{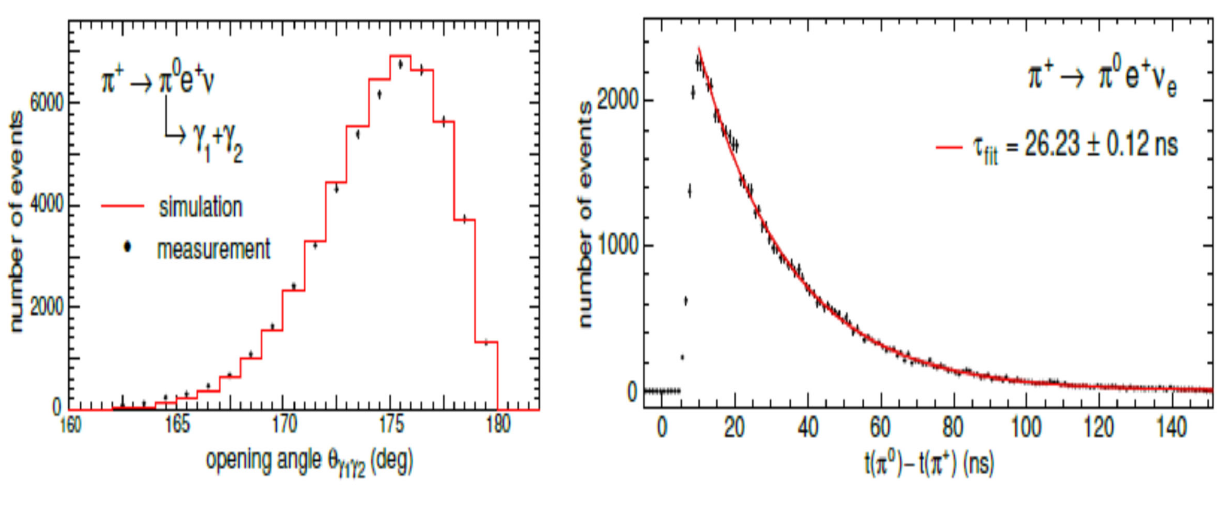}
	 \caption{(Left) Measured photon-photon opening angle from $\pi^0\rightarrow \gamma \gamma$ decay  and  Monte Carlo simulation (histogram). (Right) Decay time distribution for $\pi_{e3}$ events. Figures from ~\cite{Pocanic:2021zhy}.}
	\label{PiBeta_data}
\end{figure}

In the PiBeta experiment ~\cite{Pocanic:2003pf,Frlez:2003vg}, the $\pi_{e3}$ decay signal consisted of  two energetic, nearly back-to-back neutral particle induced clusters 
in the calorimeter, initiated by the two photons from $\pi^0$ decay.
Fig. ~\ref{PiBeta_data} shows the opening angular distribution for the photons and the time distribution of $\pi_{e3}$ events. 
 The exponential decay time distribution
 agrees  with the known pion lifetime.
To obtain the $\pi_{e3}$ branching ratio, $\pi_{e2}$  decay  i.e. Eq. ~\ref{pienu(g)Br},  was used for normalization. While the
two decays shared many of the same systematics, such as the spatial and temporal distributions
of the parent pions, and very similar acceptances, the $\pi_{e2}$ signal had a significant background
from  muon decays \mudecay. Details of the analysis and results for the $\pi_{e3}$ branching
ratio are discussed in \cite{Pocanic:2003pf}. The result is
~\cite{Pocanic:2003pf,Pocanic:2021zhy}
\begin{equation}
\begin{aligned}
\label{RpiBeta}
     R^{\rm\overline{Exp}}_{\pi_{e3}}= (1.038 \pm 0.004_{stat} \pm 0.004_{syst} \pm 0.002_{\pi_{e2}})\times 10^{-8}
     =(1.038 \pm 0.006)\times 10^{-8}.
\end{aligned}
\end{equation}
This is in  excellent agreement with the SM expectation given in Eq.~\ref{pibeta_br_sm}.

The result for $R^{\rm\overline{Exp}}_{\pi_{e3}}$ in Eq. ~\ref{RpiBeta} can be used to obtain $V_{ud}= 0.9739(28)$
based on the SM analysis given in Sec. ~\ref{pibeta_theory}.
While this is in agreement with Eq.~\ref{Vud}, the present accuracy of 0.3\%  is not competitive with the accuracy of
the fit quoted in Eq. \ref{Vud}. 
 In order for pion beta decay to provide precision on $V_{ud}$ comparable with  Eq.~\ref{Vud} an
 order-of-magnitude improvement in accuracy would be needed in the experimental measurement of 
 the branching ratio $BR(\pi^+ \to \pi^0 e^+ \nu_e)$
 along with  a factor of five improvement  in the uncertainty on $\Delta_\pi$, as indicated in Sec.~\ref{pibeta_theory}.

In related work, the SINDRUM I collaboration at PSI produced the most recent results on the pion radiative decay  process $\pi^+ \rightarrow e^+ \nu_e e^+ e^-$ ($\pi_{3e}$)~\cite{SINDRUM:1989qan}. Analysis selections resulted in the observation of 98 events including one estimated background
event, based on observing $4 \times 10^{12}$ stopped pions. 
The branching ratio  reported was
$B(\pi_{3e}) = (3.2 \pm 0.5 \pm 0.2) \times 10^{-9}$ 
where the first uncertainty is experimental and the second uncertainty is from 
theory. 
  This experiment obtained a value for the ratio $r_F = 0.7 \pm 0.5$ (see Eq.~\ref{rf}), resolving a quadratic ambiguity in earlier experiments, in agreement with the results obtained in Ref.~\cite{Bychkov:2008ws}.

\subsection{Neutral Pion Decay Experiments}\label{ExpNeutral}

\begin{figure}[H]
	\centering
\includegraphics[width=0.5\textwidth]{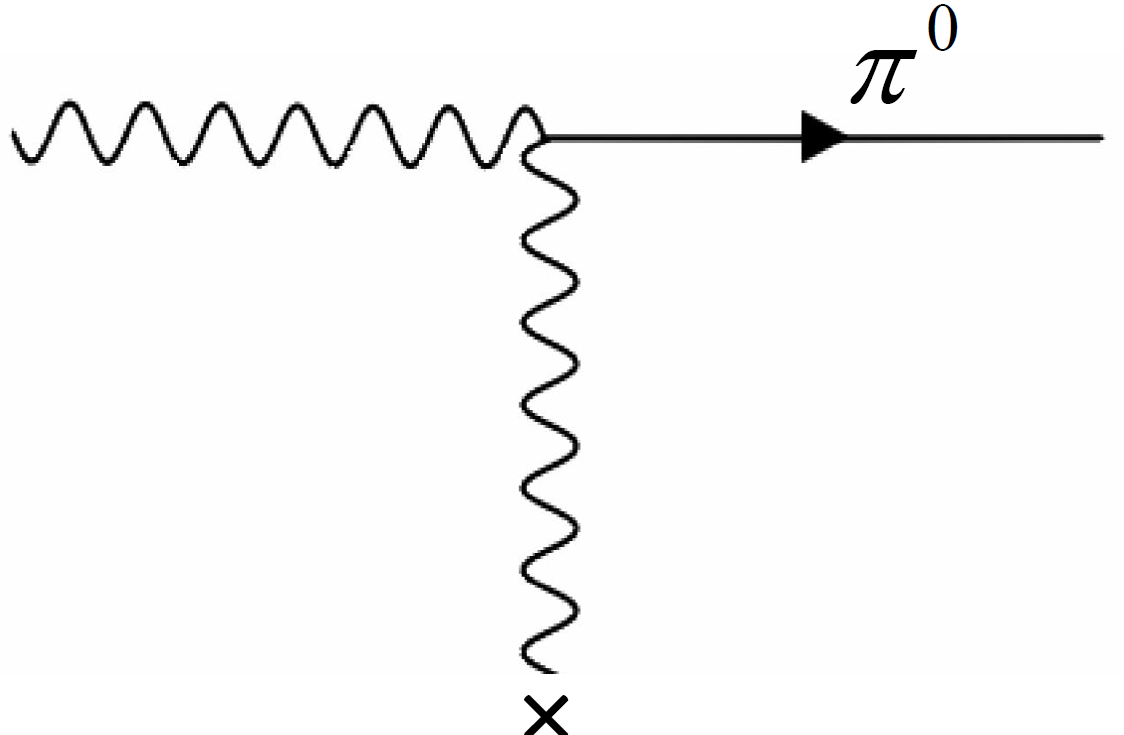}
	 \caption{Primakoff effect: ${\pi^0}$ production occurs when an incident photon intreacts with the Coulomb field of a nucleus (X). }
	\label{Primakoff}
\end{figure}

$\pi^0$ lifetime measurements have been performed using several techniques which are reviewed 
in ~\cite{Bernstein:2011bx} and listed in
~\cite{ParticleDataGroup:2024cfk}.
The
most accurate measure of the ${\pi^0}$ lifetime was reported  in 2020 by the PrimEex II experiment~\cite{PrimEx-II:2020jwd}  at the Thomas Jefferson National Accelerator Facility (JLAB). PrimEex II employed the Primakoff effect~\cite{Primakoff:1951iae}
 in which an incident photon interacts with the
Coulomb field of a nucleus to produce a ${\pi^0}$ as illustrated by the  
Feynman diagram in Fig. ~\ref{Primakoff}. PrimEx II used focused tagged photons to obtain a well controlled energy and timing source  provided by a continuous 
wave (cw) electron accelerator resulting  a  high duty cycle (i.e. high ratio of  beam on to beam off times) which reduced accidental coincidences.

\begin{figure}[H]
	\centering
\includegraphics[width=0.7\textwidth]{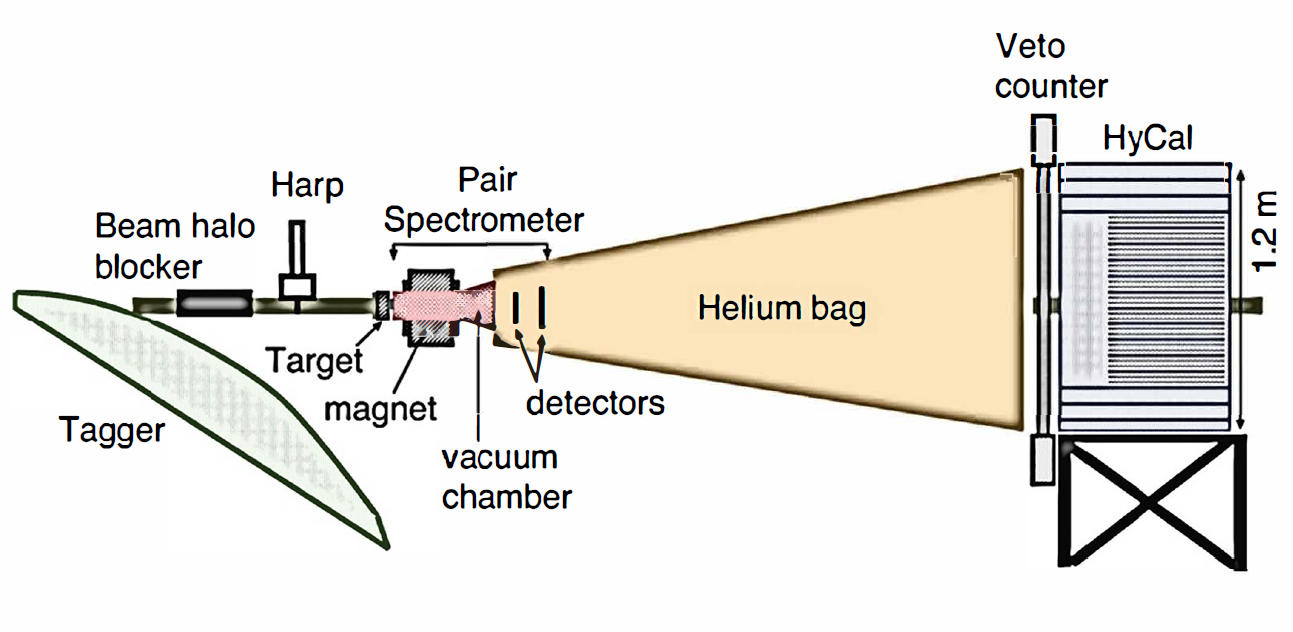}
	 \caption{PrimEx II experiment schematic\cite{PrimexIIprivate}. Not to scale. See text. }
	\label{PrimexII}
\end{figure}

A schematic diagram of the PrimEx II experiment is shown in Fig. ~\ref{PrimexII}. The  JLAB Hall B tagged photon beam facility~\cite{Sober:2000we} provides energy and angle selected photons
enhancing the 
background separation and  facilitating the determination
of the incident photon flux. 
This system can tag photon energies over a range from 20\% to 95\% of the incident electron energy, and is capable of operation with electron beam energies up to 6.1 GeV. The Tagger consists of a single dipole magnet combined with a hodoscope containing two planar arrays of plastic scintillators to detect energy-degraded electrons from a thin bremsstrahlung radiator~\cite{Sober:2000we}. 

The differential
cross sections were measured for $\pi^0$  produced by  the Primakoff effect on  $^{12}$C and  $^{28}$Si targets. The targets are located 
at the entrance of a
dipole magnet which 
deflects charged particles produced in
photon interactions
 away from the spectrometer; use of two targets helped with study of systematic uncertainties.  Photons  are  converted to $e^+ e^-$ pairs in the  spectrometer which provides angular and energy measurements. A segmented electromagnetic calorimeter (HyCal)~\cite{Gasparian:2004xa} located 7 m downstream from the targets  consists of  PbWO4 crystals. The HyCal
featured
high-resolution energy measurements and large-acceptance.  The cross
sections of Compton scattering and positron-electron
($e^{+} e^{ - }$) pair production were periodically measured for calibration purposes to monitor
 the
extracted $\pi^0$ photo-production cross sections
and their estimated systematic uncertainties.
The $e^{+} e^{ - }$ ~pairs  detected  by the pair
spectrometer  allowed continuous measurement
of the relative photon-tagging efficiencies
during the experiment. 
Two-planes of scintillation counters (Veto Counters),
located in front of HyCal, provide rejection
of charged particles, reducing background.
Elastically
produced $\pi^0$ events  were
measured  in angular bins using  kinematic constraints
and  the experimental two-photon invariant
mass spectra $M^2_{\gamma \gamma}=E_{\gamma 1} E_{\gamma 2} (1-\cos\theta)$ (with  subtraction of background contributions) were obtained by fitting. Here, $E_{\gamma 1(2)}$ is the energy measured for the first (second) gamma ray and $\theta$ is the angle between the photon directions  measured from the target using the locations of the hit elements of the HyCal.

Measurement of the ${\pi^0}$ production cross section yields the value of the  ${\pi^0}$ decay width $\Gamma_{\gamma\gamma} = \tau^{-1}_{\pi^0}$ where $\tau_{\pi^0}$ is the $\pi^0$ mean life  since the cross section of the Primakoff process is directly proportional to $\Gamma_{\gamma\gamma}$.  
The Primakoff differential cross section narrowly peaks at small angles and is given in the high
energy limit by~\cite{Gourdin:1971yq,Bernstein:2011bx} 
\begin{equation}
\begin{aligned}
\frac{d\sigma_p}{d\Omega}  = \Gamma_{\gamma\gamma}  
\frac{8 \alpha_{em} Z^2}{m^3}  \frac{k^2}{Q^2_{min}}
|F_{em}(Q)|^2 f(\frac{Q^2}{Q^2_{min}}), \\
f(t=\frac{Q^2}{Q^2_{min}})=\frac{(t-1)}{t^2}, \quad\quad  Q^2_{min}=(\frac{m^2}{2k})^2, \\
\theta_{P:max}\approx \frac{m^2}{2k^2}
\end{aligned}
\end{equation}
where $Q^2=$$-$ $ q^2$  is the negative of the square of the four-momentum transfer q to the nucleus. 
$F_{em}(Q)$ is the nuclear electromagnetic form factor~\cite{Faeldt:1972db},
Z is the atomic number of the
target nucleus, m is the mass of the $\pi^0$, k is the
energy of the incoming photon, $Q^2_ {min}$
is the minimum value for
the square of the momentum transfer, and $\theta_{P:max}$ is the angle
for which the Primakoff cross section reaches its maximum
value. The four-momentum transfer t  is dimensionless in units of $Q^2_{min}$.
The energy-independent function $f(t)$   rises rapidly at forward angles ($\theta=0$, $t = 1$)
with a peak at $t = 2$; the Primakoff cross
 decreases rapidly with photon energy~\cite{Bernstein:2011bx}  from its maximum value.

\begin{figure}[H]
	\centering
\includegraphics[width=1.\textwidth]{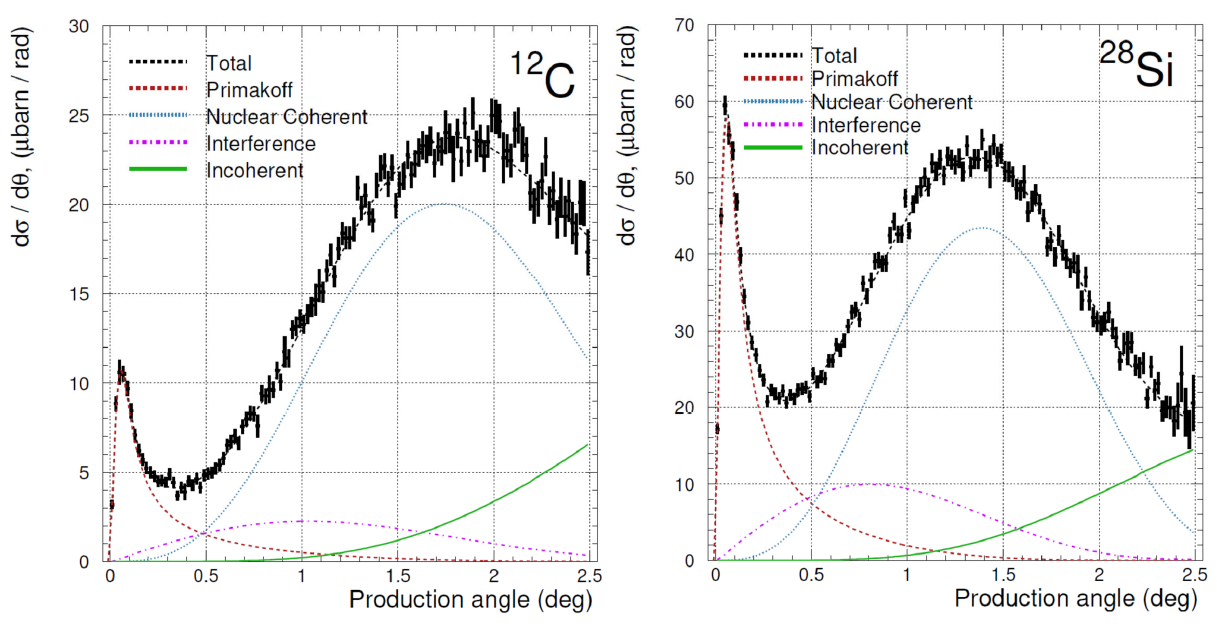}
	 \caption{ Experimental  differential cross sections\cite{PrimexIIprivate} measured by PrimEx II as a function of the $\pi^0$ production angle for $^{12}C$ (Left) and $^{28}Si$ (Right) together with the fit
results for the different physics processes; the Primakoff process shown by the red dotted line peaks at very forward angles. Error bars indicate  statistical uncertainties. }
	\label{PrimExIIdata}
\end{figure}

 For  incident photon angles from 0° to
2.5° and  energies  $E_\gamma =4.45$ to
5.30 GeV there were 83,000 $\pi^0$ events on $^{12}C$ targets and 166,000 events on $^{28}Si$ targets; the data is shown in Fig~\ref{PrimExIIdata}.
The
fit results also  shown in Fig. ~\ref{PrimExIIdata} involve  four processes that contribute
to forward $\pi^0$ production including  the Primakoff process, 
 coherent nuclear reactions,  interference between
the Primakoff and coherent  nuclear  amplitudes,
and nuclear incoherent processes.
$\Gamma_{\gamma\gamma}$ 
was extracted by
fitting the experimental differential cross sections including the  four contributing
processes  convoluted with
the angular resolution and experimental acceptances
and folded with the measured incident
photon energy spectrum. The effects of
final state interactions between the outgoing
pion and the nuclear target, and that of  photon
shadowing by nucleons  within nuclear matter was included in the theoretical cross
sections.

The results
from the two targets were  combined
to generate the final result: $\Gamma_{\gamma\gamma}=
7.798 \pm 0.056(stat) \pm 0.109(syst)$ eV, with a total
uncertainty of 1.57\%, in excellent agreement with
the SM expectation given in Sec.~\ref{pi0_theory}. The Particle  Data Group used this result in combination with previous experiments measuring $\Gamma_{\gamma\gamma}$ to find the $\pi^0$ mean life as
$\tau_{\pi^0}
=\frac{1}{\Gamma_{\gamma \gamma}}
=8.43 \pm 0.13 \times 10^{-17}$s~\cite{ParticleDataGroup:2024cfk}.

Some subdominant $\pi^0$ decay modes, such as 
the Dalitz decay $\pi^0 \to e^+ e^- \gamma$ ~\cite{Dalitz:1951aj} are also of interest.
Radiative corrections to the total and differential decay rate for
this decay mode were calculated in
~\cite{Lautrup:1971ew,Mikaelian:1972jn,Mikaelian:1972yg}. The measured branching ratio is
$BR(\pi^0 \to e^+ e^- \gamma) = (1.174 \pm 0.035)$\% ~\cite{ParticleDataGroup:2024cfk},
consistent with theoretical
calculations. Several experiments have used the reaction
$\pi^- p \to n \pi^0$ to produce the $\pi^0$ mesons for other studies, e.g. ~\cite{Samios:1961zz}. 
The double Dalitz decay mode, $\pi^0 \to 2(e^+ e^-)$ has also been measured by 
the KTeV experiment at Fermilab~\cite{KTeV:2008pev}, which obtained
$BR(\pi^0 \to 2(e^+ e^-) = (3.34 \pm 0.16) \times 10^{-5}$.
The decay $\pi^0 \to e^+ e^-$ is also of interest. Approximate calculations
of the rate for this decay have been performed in a variety of
models for the bound-state structure of the $\pi^0$
(see ~\cite{Bergstrom:1982zq,Bergstrom:1982ve} and references therein) resulting in a  lower 
bound on the branching ratio  $4.7 \times 10^{-8}$ ~\cite{Berman:1960zz}
consistent with the experimental value  $BR(\pi^0 \to e^+ e^-) = (6.46 \pm 0.33) \times 10^{-8}$ ~\cite{ParticleDataGroup:2024cfk}.

\subsection{Exotic Pion Decays}
\label{Exotics}

High-sensitivity 
pion decay experiments have been used to search for a range of exotic decays, i.e. those involving particles and interactions not included in the SM, such as heavy neutral leptons, light neutral, weakly interacting bosons, and lepton flavor violation.


The observation of neutrino oscillations showed that the neutrino weak 
eigenstates ($\nu_\ell$) are linear combinations of mass eigenstates ($\nu_i$), of the form
\beq
|\nu_\ell\rangle = \sum_i U_{\ell i}|\nu_i\rangle
\label{nuell}
\eeq
where $\ell = e, \ \mu, \ \tau$ and $U_{\ell i}$ are elements of a mixing matrix.  The sum over $i$ includes at least $1 \le i \le 3$
and may also include higher-lying mass eigenstates. From the width of the $Z$ vector boson it is known that there are three
flavors of active neutrinos; this follows from the measurement of the partial widths $\Gamma(Z \rightarrow \nu\bar\nu) \equiv \sum_{\ell=e,\mu,\tau} \Gamma(Z \rightarrow \nu_\ell \bar\nu_\ell)$
~\cite{ParticleDataGroup:2024cfk}. It was suggested in  
~\cite{Shrock:1980vy,Shrock:1981wq} to search for the possible emission of heavy neutrinos in
$\pi_{\ell 2}$ decays,
where $\ell = \mu, \ e$. These are emitted via the lepton mixing (as in Eq.~\ref{nuell}), so their contribution to the decay rate
for $\pi^+ \to \ell^+ \nu_\ell$ involves a factor 
$|U_{\ell i}|^2$, where here $i>3$ refers to the heavy neutrinos. If a heavy neutrino $\nu_i$ is emitted in $\pi^+_{\ell e}$ decay, the signature is a peak in the $dN/dE_\ell$ spectrum at an anomalously low value. This search is particularly sensitive in the case of $\pi^+_{e2}$ decay because the helicity suppression of the SM decay rate is removed for heavy neutrino emission. 
Searches for heavy neutrino emission in $\pi^+ \to e^+ \nu_e$ and $\pi^+ \to \mu^+ \nu_\mu$ decays were performed in a series of increasingly sensitive experiments at TRIUMF ~\cite{Bryman:1983cja,Azuelos:1986eg,Britton:1992xv,PIENU:2011aa,PIENU:2017wbj,PIENU:2019usb} and SIN/PSI ~\cite{Abela:1981nf,Minehart:1981fv,Daum:1987bg} (see 
~\cite{Bryman:2019bjg,Abdullahi:2022jlv} for additional references). Production and decays of heavy neutral leptons (HNL) have been searched for in many experiments, including the CERN PS191 experiment ~\cite{Bernardi:1985ny,Bernardi:1987ek}, T2K ~\cite{T2K:2019jwa}, MicroBooNE ~\cite{MicroBooNE:2019izn,MicroBooNE:2022ctm,MicroBooNE:2023eef} and NA62~~\cite{NA62:2021bji,NA62:2020mcv}. These searches have yielded very stringent upper limits on $|U_{ei}|^2$ and 
$|U_{\mu i}|^2$ for a heavy neutrino $\nu_i$; experiments on $\pi^+_{\mu 2}$ decay have obtained the limit 
$|U_{\mu 4}|^2<10^{-5}$ for a $\nu_4$ with mass in the range 16 to 33 MeV\cite{PIENU:2019usb}, while experiments on $\pi^+_{e2}$ have obtained the limit $|U_{e 4}|^2<10^{-8}$
in the $\nu_4$ mass range 60 to 120 MeV ~\cite{PIENU:2017wbj}.
Heavy neutrino emission would also change the ratio 
$R_{e/\mu}$ from its SM value, so the agreement of the measured value of this ratio with the SM prediction yields an upper bound on heavy neutrino mixings; this extends down to 
$|U_{e4}|^2 < 2 \times 10^{-7}$ for a $\nu_4$ mass of 55 MeV 
~\cite{Bryman:2019ssi,Bryman:2019bjg}. See  Refs.~~\cite{Bryman:2019bjg,Abdullahi:2022jlv,ParticleDataGroup:2024cfk} for discussion of other limits on heavy neutrinos. If the heavy neutrino were a Majorana fermion (equal to its own antiparticle), then for the masses of relevance to these experiments, there could be tension with the
non-observation of neutrinoless double beta decay, but there 
is no such tension if the heavy neutrino is a Dirac fermion (distinct from its antiparticle).  There is also a model-dependent constraint on the heavy neutrino lifetime from the requirement 
that it must decay sufficiently rapidly to avoid upsetting the 
successful predictions for H and He abundances via primordial nucleosynthesis~\cite{Dolgov:2000jw,Gelmini:2020ekg,Boyarsky:2020dzc,Bondarenko:2021cpc,Sabti:2021reh,Mastrototaro:2021wzl,Domcke:2020ety,Drewes:2024dem}.

Other exotic decay channels like $\pi^+ \rightarrow e^+ \nu_e X$\cite{PIENU:2021clt}, where X is a non-SM weakly interacting neutral boson(s), have also been searched for with null results. Similarly, reactions like $\pi^0 \rightarrow invisible$ ~\cite{NA62:2020pwi, E949:2005efl} and
$\pi^0 \rightarrow \gamma \nu \overline\nu$~\cite{NA62:2019meo} have also been sought.
 Many  theories extending the SM  predict
the existence of particles of this type. Some examples include
axion-like pseudoscalar particles (ALPs) ~\cite{Peccei:1977hh,Peccei:1977ur,Weinberg:1977ma,Wilczek:1977pj,Altmannshofer:2019yji,Altmannshofer:2022ckw}, pseudo-Nambu-Goldstone bosons (PNGBs) called Majorons associated with spontaneous breaking of total lepton number  ~\cite{Gelmini:1980re,Gelmini:1982rr,Chikashige:1980ui,Barger:1981vd}, dilatons~\cite{Girmohanta:2023tdr}, and scalar or vector particles associated with dark matter sectors ~\cite{Bertone:2004pz,Battaglieri:2017aum,Hostert:2020gou,Hostert:2020xku,Krnjaic:2022ozp}. Some limits include $BR(\pi^0 \rightarrow invisible) < 4.4\times10^{-9}$~\cite{NA62:2020pwi} and $BR(\pi^0 \rightarrow \gamma \nu \overline\nu)<1.9\times10^{-7}$~\cite{NA62:2019meo}. Searches for lepton flavor violation in $\pi^0$ decays~\cite{Bryman:1982km} have also resulted in null results including
$BR(\pi^0\rightarrow \mu^+e^-+\mu^-e^+)<3.6\times10^{-10}$~\cite{KTeV:2007cvy}.


\section{Conclusions}

Theoretical and experimental studies of pion decays have yielded important information about fundamental aspects of particle physics including high-precision tests of the $V-A$ structure of the charged weak current, lepton flavor universality,  consistency with 
the {\it ab initio} lattice QCD calculation of the pion decay constant $f_\pi$, and stringent constraints on a variety of possible physics effects beyond the Standard Model. There is an active program of research to push the experimental investigation of pion decays to even higher precision. In particular, the proposed PIONEER experiment~\cite{PIONEER:2022alm}, aims to measure the $\pi^+ \to e^+ \nu_e(\gamma)$ branching ratio ($R_{e/\mu}$) to substantially higher precision to test lepton universality and further constrain possible new physics. In later phases, PIONEER also proposes to measure pion beta decay,
$\pi^+ \to \pi^0 e^+ \nu_e$.

\label{sec:conclusions}

\begin{ack}[Acknowledgments]%
 D.B acknowledges the support of the Natural Sciences and Engineering Research Council of Canada (NSERC) grants SAPPJ-2023-00035 and SAPPJ-2024-00027.
 The work of R.S. was partially supported by the U.S. National Science Foundation via the grant NSF-PHY-22-10533. 
\end{ack}

\appendix
\section{Appendix}
\label{Appendix}

In this appendix we include some background material for the non-expert reader, including basic formulas on Dirac algebra, decay rates, and properties and interactions of pions. 

\subsection{Dirac matrices}

As noted in the text, we use the $(+,-,-,-)$ metric, i.e., the metric tensor is $g_{\mu\nu}={\rm diag}(1,-1,-1,-1)$. The Dirac gamma matrices satisfy 
\beq
\{ \gamma_\mu,\gamma_\nu \} = 2g_{\mu\nu} \ ,
\label{gamma_anticommutator}
\eeq
where $[A,B]=AB-BA$ and $\{A,B\} = AB+BA$ are the commutator and anti-commutator of matrices $A$ and $B$. Our conventions for Dirac matrices follow those of ~\cite{Itzykson:1980rh}. In chiral projections of fermion fields the $\gamma_5$ matrix enters, satisfying the defining property 
\beq
\{\gamma_5,\gamma_\mu\}=0,
\label{gamma5_anticommutator}
\eeq
where $\mu = 0, \ 1, \ 2$, or 3. Furthermore, 
\beq
\gamma_5^2 = 1 \ .
\label{gammasq}
\eeq
We take 
\beq
\gamma_5 = i\gamma^0\gamma^1\gamma^2\gamma^3.
\label{gamma_5}
\eeq
The operators 
\beq
P_R = \frac{1}{2}(1+\gamma_5) \ , \quad 
P_L = \frac{1}{2}(1-\gamma_5)
\label{plpr}
\eeq
are chiral projection operators satisfying 
\beq
P_R^2=P_R, \quad P_L^2=P_L, \quad P_LP_R = P_RP_L = 0, 
\label{plpr_projectors}
\eeq
and
\beq
P_L+P_R=1 \ ,
\label{plpr_completeness}
\eeq
where here $1 = {\rm diag}(1,1,1,1)$ is the identity matrix in Dirac space. 
Thus, the left- and right-handed chiral components of a fermion field $\psi$ are given by
\beq
\psi_L = P_L \psi \ , \quad \psi_R = P_R \psi \ .
\label{chiral_components}
\eeq

\subsection{Some Basic Kinematic Formulas}

Consider the decay of a spin-0 parent particle $A$ with mass $m_A$ and four-momentum $p_A$, satisfying $p_A^2=m_A^2$, to a final state $f.s.$ consisting of $n$ particles $b_i$, $i=1,...n$
with four-momenta $p_i$. Denoting the decay amplitude as 
${\cal M}(A \to b_1 + ... + b_n)$, the total decay rate is
\beq
\Gamma(A \to f.s.) = \frac{S}{2m_A} \, \int dR_n |{\cal M}|^2 \ , 
\label{gamma_general}
\eeq
where a sum over polarizations of final-state particles is understood;
$S$ is a symmetry factor to take account of possible identical particles in
the final state; and the integration over the $n$-body final-state phase space
is given by
\beq
\int dR_n = \frac{1}{(2\pi)^{3n-4}} \,
\int \Big [ \prod_{i=1}^n
\frac{d^3 p_i}{2E_i} \Big ] \, \delta^4 \Big ( p_A-(\sum_{i=1}^n p_i) \Big ) \ .
\label{phase_space_integral}
\eeq
In particular, for the case of two identical particles in the final state, as is relevant for $\pi^0 \to \gamma\gamma$ decay, the symmetry factor is $S=1/2!$.  For two-body decays, the three-momenta of the two final-state particles have fixed magnitudes, so the two-body phase space integration is independent of 
$|{\cal M}|$, which thus factors out of the integral. Denoting 
\beq
R_n \equiv \int dR_n \ ,
\label{rn}
\eeq
one has the result
\beq
R_2 = \frac{1}{8\pi} \, [\lambda(1,\delta_1,\delta_2)]^{1/2}
\label{r2}
\eeq
where
\beq
\lambda(x,y,z) = x^2+y^2+z^2-2(xy+yz+zx)
\label{lam}
\eeq
and
\beq
\delta_i = \Big (\frac{m_i}{m_A} \Big )^2 \ .
\label{deltai}
\eeq
In this case of a 2-body final state, if one of the final-state particles has zero or negligibly small value, then the $\lambda$ function simplifies as $\lambda(x,y,0)=(x-y)^2$.

\subsection{Further Details of Pion Properties and Interactions}

Pions transform as an $I=1$ (isovector) representation under the strong isospin SU(2) symmetry group, with the respective three isospin states $|I,I_3\rangle = |1,\pm 1\rangle$, and $|1,0\rangle$. We recall that the group SU($N$) is the group of unitary $N \times N$ matrices with unit determinant.  An element $U$ of the group SU($N$) can be written as 
\beq
U = \exp(i {\vec T} \cdot {\vec a})
\label{u}
\eeq
where ${\vec a}$ is a real $p$-dimensional vector and ${\vec T} = (T_1,...,T_p)$, $p=N^2-1$, with $T_j$ being the generators of the associated Lie algegra, satisfying the commutation relation 
\beq
[T_j,T_k] = i c_{jk\ell} T_\ell
\label{sun_commutation}
\eeq
and the normalization condition 
\beq
{\rm Tr}(T_j T_k) = \frac{1}{2}\delta_{jk}.
\label{normalization}
\eeq
In Eq. (\ref{sun_commutation}), the $c_{jk \ell}$ are the 
structure constants of the Lie algebra. 
In particular, for SU(2), the generators are $T_j = (1/2)\tau_j$, $j=1,2,3$, where (in a standard convention) 
the $\tau_j$ are the Pauli matrices 
\beq
\tau_1 = \left( \begin{array}{cc}
        0 & 1 \\
        1 & 0 \end{array} \right ), 
\quad 
\tau_2 = \left ( \begin{array}{cc}
         0 & -i \\
         i &  0 \end{array} \right ), 
\quad
\tau_3 = \left ( \begin{array}{cc} 
         1 & 0 \\
         0 & -1 \end{array} \right ) \ .
         \eeq
In addition to the spin-parity $J^P=0^-$, the pions have $G$-parity $G=-1$, where here $G=C(-1)^I$, and equivalently, the $C$-parity of the self-conjugate state $\pi^0$ is $C=+1$.

The pions are the lightest hadrons, and the smallness of their masses is understood as a consequence of the fact that they are approximate Nambu-Goldstone bosons (NGBs) resulting from the spontaneous breaking of a global 
${\rm SU}(2)_L \otimes {\rm SU}(2)_R$ symmetry to the diagonal (vectorial isospin) subgroup ~\cite{Nambu:1960tm,Goldstone:1961eq,Goldstone:1962es}.
This global chiral symmetry is not exact because of the small but nonzero ``current-quark" masses of the $u$ and $d$ quarks and the presence of electromagnetism.

In the Lagrangian for a general scalar or pseudoscalar particle ($s$), ${\cal L} = (1/2)(\partial_\mu s)^2 - (1/2)m_s^2 s^2 - V(s)$,
where internal indices are suppressed and $V$ contains terms higher
than quadratic and also interactions, the mass of the particle occurs
as a square. Taking this into account, the $\pi^\pm$ and $\pi^0$
squared masses exhibit a striking property, namely that they are 
much smaller than the typical hadronic scale of $\sim 1$ GeV:
\beq
\frac{(m_{\pi^\pm})^2}{1 \ {\rm GeV}^2 } = 0.019, \ \       \quad\quad 
\frac{(m_{\pi^0})^2}{1 \ {\rm GeV}^2 } = 0.018.
\label{pimsq}
\eeq
The Nambu-Goldstone theorem ~\cite{Nambu:1960tm,Goldstone:1961eq,Goldstone:1962es}
states that if a theory is invariant under a global symmetry but the vacuum and hence the particle states of the theory are not invariant, then the theory also contains massless, spinless bosons, called Nambu-Goldstone bosons. The Standard Model (SM) has the gauge group 
\beq
G_{\rm SM} = {\rm SU}(3)_c \otimes {\rm SU}(2)_L \otimes {\rm U}(1)_Y \ .
\label{gsm}
\eeq
Since the $u$ and $d$ quark masses $m_u$ and $m_d$ are only a few MeV,
much smaller than the QCD scale $\Lambda_{QCD} \simeq 250$ MeV, it
follows that, to a good approximation, the color SU(3)$_c$ part of the
SM Lagrangian for the $u$ and $d$ quarks is (suppressing color
SU(3)$_c$ indices)
\beq
    {\cal L} = \bar \psi \gamma \cdot D \psi 
    = \bar \psi_L \gamma \cdot D \psi_L + \bar \psi_R \gamma \cdot D \psi_R 
    \label{udterm}
\eeq
with no $u$ or $d$ quark mass terms, 
where $\psi = {u \choose d}$ is the weak isodoublet of $u$ and $d$
quarks; $\psi_L$ and $\psi_R$ are the left- and right-handed chiral
components of $\psi$ defined by $\psi_L = P_L\psi$ and $\psi_R = P_R\psi$,
where $P_{R,L} = (1/2)(1 \pm \gamma_5)$ are chiral projection operators
in Dirac space; $\gamma \cdot D = \gamma_\mu D^\mu$, $\gamma_\mu$ are
the Dirac matrices; and $D^\mu$ is the SU(3)$_c$ gauge-covariant
derivative. Here we neglect electroweak interactions, since they are
weak compared with color interactions, and also for simplicity we
consider just the first generation of SM fermions.  With the indicated neglect of $u$ and $d$ mass term, this Lagrangian is invariant under a classical
global ${\rm U}(2)_L \otimes {\rm U}(2)_R$ symmetry, where the
elements of the U(2)$_L$ (resp. U(2)$_R$) group operate on $\psi_L$
(resp. $\psi_R$). One can express this group as ${\rm U}(2)_L \otimes
{\rm U}(2)_R = {\rm SU}(2)_L \otimes {\rm SU}(2)_R \otimes {\rm
  U}(1)_V \otimes {\rm U}(1)_A$, where U(1)$_V$ is the vector quark number and
U(1)$_A$ is the axial quark number. The U(1)$_A$ classical symmetry is
anomalous, being broken at a non-perturbative level by QCD instantons ~\cite{tHooft:1976rip,tHooft:1976snw,Callan:1976je,Jackiw:1976pf,tHooft:1986ooh}, so the
actual global symmetry group here, with $m_{u,d}=0$ and electroweak
interactions neglected, is
\beq
G_{gb} = {\rm SU}(2)_L \otimes {\rm SU}(2)_R \otimes {\rm U}(1)_V.
\label{ggb}
\eeq
Nonperturbative strong QCD interactions produce a nonzero bilinear
quark condensate $\langle \bar \psi \psi \rangle$, dynamically
breaking
\beq
G_{gb} \to {\rm SU}(2)_V \otimes {\rm U}(1)_V  \ , 
\label{ggb_breaking}
\eeq
where SU(2)$_V$ is the global SU(2) isospin symmetry group and U(1)$_V$ is
the quark number U(1) symmetry group, and the subscript $V$ indicates that the elements of these groups act vectorially, i.e., in the same way on left- and
right-handed chiral components of $\psi$. By the Nambu-Goldstone theorem, this
spontaneous symmetry breaking of the ${\rm SU}(2)_L \otimes {\rm SU}(2)_R$
symmetry to the vectorial isospin group SU(2)$_V$ produces   
three resultant massless spin-0 NGBs, which are the $\pi^\pm$ and $\pi^0$.
To complete the description, one next takes account of the two main features
of nature that explicitly break $G_{gb}$, namely (i) the fact that $m_u$ and
$m_d$ are not exactly zero, although $m_{u,d} \ll \Lambda_{QCD}$, and
(ii) the fact that the electromagnetic and weak interactions are nonzero.
(The importance of electromagnetic interactions is evident, since the
$T_\pm = (1/2)\tau_\pm$ generators of SU(2)$_L$ interchange $u_L$ and $d_L$,
and similarly for SU(2)$_R$, which breaks electric charge conservation.) 
In the presence of explicit symmetry-breaking terms, the NGBs are not exactly
massless, but the smalless of both (i) and (ii) relative to the color QCD
interactions at the hadronic scale mean that the physical pions have small
masses on this hadronic scale.  This theoretical background is the
foundation for the class of low energy effective Lagrangians called
chiral Lagrangians, which have been used extensively for approximate
calculations of pion properties. 
(For references to the
original literature see, e.g., ~\cite{Donoghue:1992dd}.)
One important result is that taking into
account nonzero quark masses but still neglecting electroweak interactions,
one derives an approximate expression for the pion mass, namely
\beq
m_\pi^2 = \frac{2(m_u+m_d)}{f_\pi^2} \langle \bar \psi \psi \rangle
\label{msq_rel}
\eeq
where $f_\pi$ is the pion decay constant (see Eqs. (\ref{fpirel}) and (\ref{fpi_lattice})), and
$\langle \bar \psi \psi \rangle = \langle \bar u u + \bar d d \rangle$ is the quark condensate. (In the literature,
one commonly finds definitions of $f_\pi$ that differ from this by factors
of $1/\sqrt{2}$ or $\sqrt{2}$.) Including electromagnetic effects, one expects
on general grounds that $m_{\pi^+}$ should be larger than $m_{\pi^0}$ by
the Coulombic energy due to its charge. A rough estimate would thus be
$m_{\pi^\pm} - m_{\pi^0} \sim \kappa \alpha_{em}/R_\pi \sim
\kappa \alpha_{em} \Lambda_{QCD}$, where $R_\pi$ is the radius of the pion,
and $\kappa$ is a pure number of O(1), and this
agrees approximately with the actual mass difference, 
$\Delta_\pi$.

Concerning pion weak interactions, the conserved vector current (CVC) property states that if one writes the electromagnetic current as an isovector part and an isoscalar part, 
$J_{em,\mu} = J_{{\rm isovector},\mu} + J_{{\rm isoscalar},\mu}$, then the isovector part is in the same isospin multiplet as the charge-raising and charge-lowering vectorial parts of the weak charged current, and therefore (in the limit of exact isospin symmetry) the latter are conserved, just as  is $J_{em,\mu}$.

In Sec.~\ref{pi0_theory} on $\pi^0$ decays, mention is made of the
baryon spectroscopy argument for $N_c=3$ colors. We briefly
review this argument. The lowest-lying color-singlet bound states of quarks involving $u$ and $d$ form a 
spin 1/2 (in units of $\hbar$), isospin $I=1/2$ state, the 
nucleon isodoublet, $(uud)=p$ and $(udd)=n$, and a spin 3/2, $I=3/2$ state, the $\Delta$, with mass 1.3 GeV, comprised of $(uuu)=\Delta^{++}$,
$(uud)=\Delta^+$, $(udd)=\Delta^0$, and $(ddd)=\Delta^-$. 
In a quark-model context, the wave function for the 
$\Delta$ can be written as 
\beq
\psi = \psi_{\rm space} \,\psi_{\rm spin} \, \psi_{\rm flavor} \, \psi_{\rm color}.
\eeq
Since the $\Delta$ has half-integral spin, it is a fermion, so 
its wave function must be antisymmetric under interchange of constituent quarks.  Since it is a ground-state baryon, its 
spatial wave function  $\psi_{\rm space}$ is symmetric under this interchange. Since the $\Delta$ has spin 3/2, all three quark spins are aligned, and $\psi_{\rm spin}$ is symmetric under interchange of quarks. The flavor wave function $\psi_{\rm flavor}$ is symmetric under interchange of $u$ and $d$ flavors. The color wave function $\psi_{\rm color}$ provides the necessary antisymmetrization, since it is of the form $\epsilon_{abc}q^q q^b q^c$, where $\epsilon_{ijk}$ is the totally antisymmetric tensor density on three indices.  The property that this color wavefunction consists of a (color-singlet) product of three quarks implies that there are $N_c=3$ colors. Note that the fact that there are two quarks with masses much less than the QCD scale, namely $u$ and $d$, and one quark, $s$, with mass comparable to this scale, leading to approximate flavor SU(3) symmetry, is unrelated to the SU(3) of color.



\bibliographystyle{Numbered-Style} 
\bibliography{main.bib}

\end{document}